\documentclass{bmcart}
\pdfoutput=1
\usepackage{graphicx}
\usepackage{subfig}
\usepackage{color}
\usepackage{algpseudocode}
\usepackage{algorithm}
\usepackage{stfloats}
\usepackage{tabularx}
\usepackage{multirow}
\usepackage{rotating}
\usepackage[hyperfootnotes=false]{hyperref}
\usepackage{endnotes}
\let\footnote=\endnote
\usepackage[utf8]{inputenc} 
\usepackage[width=.96\textwidth]{caption}

\newcommand{\charm}{\textsc{Charm++}}
\newcommand{\changa}{{\sc ChaNGa}}
\newcommand{\lbmgr}{\textsc{\emph{LBManager}}}

\newcommand{\pe}{PE}
\newcommand{\pes}{PEs}
\newcommand{\tpr}{tree piece proxy}
\newcommand{\tprs}{tree piece proxies}

\definecolor{MyBlue}{rgb}{1,0,0}

\definecolor{MyOrange}{rgb}{1,.5,0}
\newcommand{\comment}[1]{}

\begin{document}

\begin{frontmatter}
\begin{fmbox}

\dochead{Research}

\title{Adaptive Techniques for Clustered N-Body Cosmological Simulations}

\author[
   addressref={aff1},                   
   corref={aff1},                       
   email={gplkrsh2@illinois.edu}   
]{\inits{HM}\fnm{Harshitha} \snm{Menon}}
\author[
   addressref={aff1},
   email={wesolwsk@illinois.edu}
]{\inits{LW}\fnm{Lukasz} \snm{Wesolowski}}
\author[
   addressref={aff1},
   email={gzheng@illinois.edu}
]{\inits{GZ}\fnm{Gengbin} \snm{Zheng}}
\author[
   addressref={aff1},
   email={pjetley2@illinois.edu}
]{\inits{PJ}\fnm{Pritish} \snm{Jetley}}
\author[
   addressref={aff1},
   email={kale@illinois.edu}
]{\inits{LVK}\fnm{Laxmikant} \snm{Kale}}
\author[
   addressref={aff2},
   email={trq@astro.washington.edu}
]{\inits{TRQ}\fnm{Thomas} \snm{Quinn}}
\author[
   addressref={aff2},
   email={fabiog@astro.washington.edu}
]{\inits{FG}\fnm{Fabio} \snm{Governato}}

\address[id=aff1]{
  \orgname{Department of Computer Science, University of Illinois at Urbana-Champaign}, 
  \cny{USA}                                    
}
\address[id=aff2]{%
  \orgname{Department of Astronomy, University of Washington},
  \cny{USA}
}

\end{fmbox}

\begin{abstractbox}

\begin{abstract} 
\changa{} is an N-body cosmology simulation application implemented using
\charm{}. In this paper, we present the parallel design of \changa{} and address many
challenges arising due to the high dynamic ranges of clustered datasets. We
focus on optimizations based on adaptive techniques for scaling to more than
$128K$ cores.  We  demonstrate strong scaling on up to $512K$ cores of Blue
Waters evolving $12$ and $24$ billion particles. We also show strong scaling of
highly clustered dataset on up to $128K$ cores.

\end{abstract}



\end{abstractbox}

\end{frontmatter}

\section{Introduction}
\label{sect:intro}
Simulating the process of cosmological structure formation with enough
resolution to determine galaxy morphologies
requires an enormous dynamic range in space and time. Star formation
(SF) is concentrated in dense gas clouds the size of just a few parsecs,
while the assembly of galaxies happens over billion of years,
driven by large scale structures extending over megaparsecs.

Constraints on cosmology are tightest on scales of tens of
megaparsecs and larger due to observations of the Cosmic Microwave
Background, giving us detailed initial conditions \cite{Planck2013};
however our knowledge of the non--linear evolution of the Universe and
of the properties of galaxies is still imperfect, as the detailed
properties of Dark Matter \cite{brooks14r} and of SF \cite{review14}
remain only partially understood.  On the other hand, simulations of
large volumes of the Universe \cite{davis85,eagle14}, and of
individual galaxies at high resolution \cite{eris11,hopkins13} have
been fundamental in putting the standard hierarchical, Cold Dark
Matter dominated model, on a robust footing \cite{frenkwhite12}.
Further understanding requires numerical simulations of increasing
dynamical range, mass and spatial resolution and physical complexity,
providing a powerful incentive to develop ever more sophisticated parallel codes
\cite{vogelsberger12,AGORA}.

Scaling such codes to large processor count requires overcoming not
only spatial resolution challenges, but also large ranges in timescales.
Ignoring these multiple timescales leads to the use of
uniformly small time steps, and thus a lot more work, although it is
easier to parallelize. Using different time steps for different
particles is potentially more efficient but leads to an algorithm that
is significantly harder to parallelize effectively.

This paper presents the design of \changa{}, a parallel {\it
n-body$+$SPH} cosmology simulation program for the simulations of
astrophysical systems on a wide range of spatial and time scales.
Most of the physical modules of \changa{} have been imported from the
well established tree$+$SPH code {\sc GASOLINE} and we refer the
readers to the existing literature \cite{Wadsley04,wadsley08,G14} for
more details.


In this paper we focus on the optimizations
implemented in it to scale to large numbers of processors, and to
deal with the challenges brought on by the high dynamic ranges of
clustered datasets. However, we will begin with an overview of the
field and place the approach taken by \changa{} in the context of
published material. We then briefly summarize some specific features
of \changa{} (some imported from {\sc GASOLINE}), including force
softening, smooth particle hydrodynamics, star formation, and
multi-stepping. The parallel design of \changa{}, based on
over-decomposition of work, allowing a parallel run-time system to
dynamically balance it, is presented next, along with descriptions of
the phases of the computation. To set the context, and a baseline, for the
optimizations presented, we first describe the single-stepping
performance on relatively uniform data-sets. The clustered data-sets
are then introduced, and a series of performance challenges along with
strategies and optimizations developed to overcome them are
described. These are accompanied by detailed performance analysis
using the Projections performance visualization tool~\cite{Projections}. As of
Spring 2014 our performance evaluation runs demonstrate good speedups
to over 131,000 processor cores on NCSA's Blue Waters and up to a 3x
speedup over the single-stepping algorithm \footnote{A public version of \changa{} is publicly
available at {\tt http://hpcc.astro.washington.edu/tools/changa.html}}.

\section{Current State of the Art}
\label{sect:related}
Because of the computational challenge and the non-trivial algorithms
involved, cosmological N-body simulations have been an extensively
studied topic over the years.  In order to frame our work in \changa{},
we review some of the recent successes in scaling
cosmological simulations on the current generations of
supercomputers.  However, direct comparison of the absolute
performance among different codes is difficult.  Different choices of
accuracy criteria for the force evaluations and the time integration
will have a big impact on performance, and the choices for these
criteria will be determined by the various scientific goals of the
simulation.  For example, understanding the development of structures
at very high redshift (e.g. \cite{Ishiyama2012}) will present different
parameter and algorithm choices than simulations that model the
observations of current large scale structure (e.g. \cite{Habib2013}).

2HOT\cite{Warren2013} is an improved version of the HOT code which has been developed over the past two decades.  It uses an Oct-tree for gravity and its
gravity algorithm is very similar to that of \changa{}.  This code
demonstrates near perfect strong scaling out to 262 thousand cores on
Jaguar with a 128 billion particle simulation, implying 500,000
particles per core at the largest core count.  The actual size of the
scaling simulation (in Gigaparsecs) was not reported, but can be
presumed to be a box of order 1 Gigaparsec based on the other
simulations presented in \cite{Warren2013}.  HOT2 does implement a
multi-step time-stepping algorithm, although it is not clear whether
particles have individual time steps, and performance for the multi-step
method was not presented.

The HACC~\cite{Habib2013} framework scales to millions of cores on a
diverse set of
architectures. It uses a modified TreePM algorithm: an FFT based
particle-mesh on the large scales, a tree algorithm on intermediate
scales and particle-particle on the smallest scales.
HACC has been demonstrated to scale with near perfect parallel
efficiency out to 16384 nodes on Titan with 1.1 trillion particles,
and out to 1.6 million cores on Sequoia with 3.6 trillion particles.
These are weak scaling results, typically with millions of particles
per core.  They also demonstrated strong scaling out to 8182 nodes on
Titan and 16384 cores on Sequoia.

The GreeM code \cite{Ishiyama2012} demonstrates scaling of a trillion particle
simulation to 82944 nodes (663522 cores) of the K computer, implying
1.5 million particles per core.  This code also uses a TreePM
algorithm with a hand-optimized particle force loop and a novel method to
parallelize the FFT.  They report that despite the new parallelization
method, the FFT remains the bottleneck in their TreePM code.  They
also employ a multi-step method that splits the PM and particle forces,
but the particles do not have individual time steps.

The Gadget-3 TreePM code was used to perform a large scale structure,
DM-only simulation (the ``Millenium XXL'') on 12288 cores using 303
billion particles \cite{Angulo2012}.  With over 16 million particles
per core, special effort was needed to optimize the memory usage of
the code because the simulation was limited by memory resources.

Most of these cosmological N-body codes with published performance
data scale to millions of cores with almost perfect parallel
efficiency, given very large problem size (typically trillions of
particles). However, it becomes even more challenging to simulate a
relatively smaller problem size with higher resolution using large
number of cores. This is due to the fact that the distribution of
clusters of particles in the simulated system tends to become more
non-uniform as resolution increases, leading to load imbalance and
making it hard to scale.  The addition of hydrodynamics and cooling
only exacerbates this problem.  Recent projects that coupled gravity
with hydrodynamics in galaxy formation simulations and scaled past a
few thousand cores include EAGLE and Illustris. The codes used
(GADGET-3 and AREPO) share many of the features of \changa{} that are
necessary for galaxy formation, including individual time steps for
particles, gas dynamics, and star formation/feedback prescriptions
\cite{eagle14,illustris14}.   While some codes handle non-uniform
distributions well (e.g. Gadget-3) they have not been shown 
yet to scale to large (100,000 cores or greater) core counts.
Hence, to our knowledge,  \changa{} is the first code to 
explicitly tackle both the uniform and highly clustered simulations
 with extremely large scaling. This is achieved by several techniques
including multi-stepping and large scale dynamic load balancing
described in Section~\ref{sect:multistep}.


\section{ChaNGa}
\label{sect:changaintro}
The N-body/Smooth Particle Hydrodynamics (SPH)
code \changa{} \cite{2007_ChaNGaScaling,2009ChaNGaGPU}, is an application
implemented using Charm++.  \changa{} includes a number of features
appropriate for the simulation of cosmological structure formation,
including high force accuracy, periodic boundary conditions, evolution
in comoving coordinates, adaptive time-stepping, equation of state
solvers and subgrid recipes for star formation and supernovae
feedback.  The code is also being compared with similar codes in the
AGORA comparison project \cite{AGORA14}.  Cosmology research based on
\changa{} includes modeling the impact of a dwarf galaxy on the Milky
Way \cite{Purcell11}, modeling the intracluster gas properties in
merging galaxy clusters \cite{Ruan13} and distinguishing the role of
Warm Dark Matter in dwarf galaxy formation and structure 
 \cite{G14}.  In this section we describe the features of \changa{},
particularly as they relate to cosmological structure formation.  In
addition to the physics features described below, \changa{} has a number
of usability features required for pushing a large simulation through
a production system, such as the ability to efficiently checkpoint and
restart on a different number of processors.

\subsection{Gravitational Force Calculation}

The gravitational force calculation is based on a modified version of
the classic Barnes-Hut algorithm \cite{barnes86}.  Details of our
modifications are described in section \ref{sect:parapproach}, and
many of our optimizations are taken from PKDGRAV \cite{Stadel01}, upon
which our gravity calculation is based.  As in PKDGRAV, we choose to
expand to hexadecapole order the multipoles used for evaluating the far
field due to a mass distribution within a node.  For the force
accuracies required for cosmological simulations, better than 1
percent \cite{Power03,Reed03}, this higher order expansion is more
efficient \cite{CharmAppsChaNGa}.

\subsection{Force Softening}

When simulating dark matter and stars, the goal is to understand the
evolution of a smooth distribution function that closely approaches a
Boltzmann collisionless fluid. As the N-body code is sampling this
distribution using particles, a more accurate representation of the
underlying mass distribution is obtained if the particles are not
treated as point masses, but instead have their potential
softened \cite{Dehnen01}.  Softened forces are also of practical use
since they limit the magnitude of the inter-particle force. Typically,
the softening length is set at the inter-particle separation at the
center of DM (Dark Matter) halos \cite{Power03}.

Calculating the non-Newtonian forces introduced by softening adds a
complication to the multipole calculation: Newtonian forces have
symmetries which greatly reduce the complexity of higher order
multipoles, and the number of components of the multipole moments that
need to be stored.  \changa{} implements softening using a cubic spline
kernel, whose compact support means this complexity is not needed
beyond a specified separation (convergence with Newtonian gravity
is formally achieved at two softening lengths).  Furthermore, rather
than evaluating the more complex multipoles when softening is
involved, \changa{} evaluates all forces involving softening using only
the monopole moments, using a stricter opening criterion to maintain
accuracy.

\subsection{Periodic boundary conditions}

In order to efficiently and accurately simulate a portion of an
infinite Universe, we perform the calculation assuming periodic
boundary conditions.  Because of the long range nature of gravity, the
sum over the infinite number of periodic replicas converges very
slowly.  \changa{} accelerates this convergence using Ewald
summation \cite{Ewald21}, implemented similarly to \cite{Ding92} as more
fully described in \cite{Stadel01}.  This technique has the advantage
that the non-periodic force calculated from the tree-walk is not
modified, and therefore is simple and fast.

\subsection{Multi-stepping}

In order to efficiently handle the wide range of timescales in a
non-uniform cosmological simulation, \changa{} allows each particle to
have its own time step.  In order to amortize overheads associated with
the force calculation, such as tree building, the time steps are
restricted to be power-of-two subdivisions of the base time step.
Details of this scheme, including how to integrate the equations of
motion in coordinates that follow the expansion of the Universe, are
described in \cite{Quinn97}.

\subsection{Smooth Particle Hydrodynamics}

\begin{figure*}[t]
\includegraphics[width=0.96\textwidth]{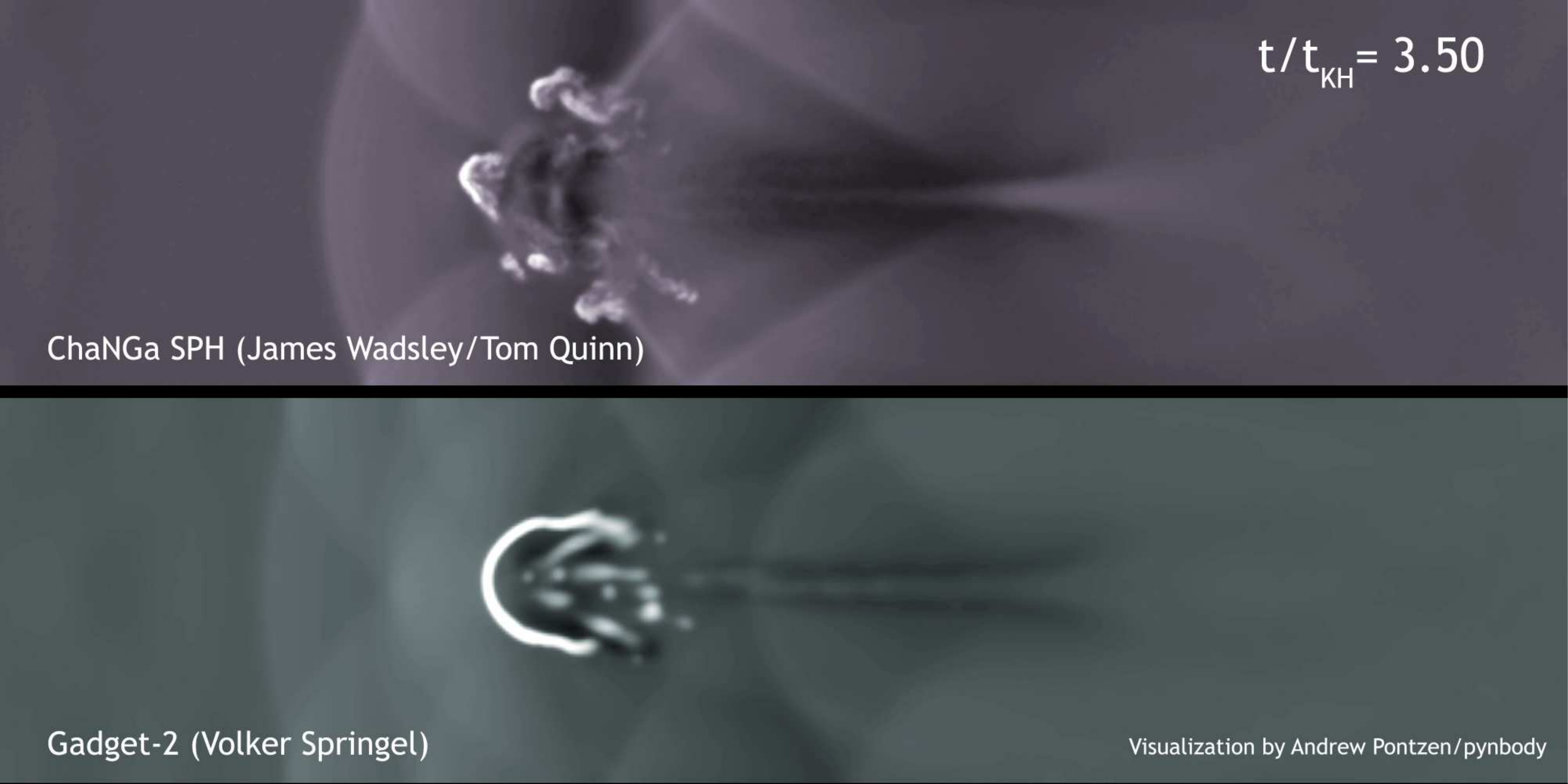}
\caption{ {\small\baselineskip=0.2cm{{\sc {Updated
          modeling of gas physics in \changa{}: the qualitative`blob'test.}}  Central
      density slices of the time evolution of a high density cloud in
      pressure equilibrium in a wind.  Time is in units of the
      Kelvin-Helmholtz growth time. \changa{} (top) vs GADGET-2
      (bottom). The color density map shows how with the new SPH
      formulation of pressure gradients, artificial surface tension is
      suppressed and instabilities rapidly mix the ``blob'' with the
      surrounding medium, while poor handling of contact
      discontinuities preserve the blob in the now obsolete SPH
      implementation of GADGET-2 (GADGET-2 was originally introduced in
      2001, see \cite{gadget01}). We have verified that \changa{} gives
      results quite similar to  alternative hydro codes, as the
      adaptive mesh refinement code ENZO \cite{collins10}.  This
      figure was produced with Pynbody \cite{pynbody}.}}}
\label{fig:blob_changa}
\end{figure*}

Despite being a small fraction of the energy density of the Universe,
baryons play a significant role in the evolution of structure.  Not
only are they the means by which we can measure structure (e.g. via
star light), they can also directly influence the structure of the
dark matter via gravitational coupling \cite{PontzenGovernato12}.
Therefore following the physics of the baryonic gas is essential for
accurate modeling of structure formation.  \changa{} uses Smooth Particle
Hydrodynamics (SPH) to solve the Euler equations with an
implementation that closely follows \cite{Wadsley04}.  Since SPH is
based on particles, implementing it is a natural extension of the
algorithms to calculate gravity on a set of particles.  In particular,
the tree structure used for the Barnes-Hut algorithm is used to find
the near neighbor particles needed for the SPH kernel sums.  SPH is
also relatively communication intensive compared to gravity, so we
utilize the Charm++ runtime system to adaptively overlap the
communication latencies from SPH with the floating point operations
needed by gravity. The current implementation of SPH in \changa{}
closely follows techniques already published by independent groups and
includes an updated treatment of entropy and thermal diffusion
\cite{wadsley08,shen10}, pressure gradients\footnote{Using a
geometric density mean in the SPH force expression:
$(P_i+P_j)/(\rho_i\rho_j)$ in place of $P_i/\rho_i^2+P_j/\rho_j^2$
where $P_i$ and $\rho_i$ are particle pressures and densities
respectively.}  and timestepping \cite{durier12}. This last features
ensures that sudden changes in the particle internal energy,
e.g. caused by feedback, are captured and propagated to neighboring
particles by shortening their time step. These improvements leads to a
marked improvements in the treatment of shocks (as in the Sedov-Taylor
blastwave test), and cold-hot gas instabilities.  A qualitative
example is shown in figure~\ref{fig:blob_changa}, where the classic ``blob'' test compares
\changa{} with GADGET-2.


As this paper focuses specifically on the scaling performance of \changa{} we
refer to existing works \cite{Wadsley04,G14} and Wadsley et al. (in prep.)
for SPH tests of this implementation.

\subsection{Star Formation and Feedback}
 
Again, a necessary component of simulating structure formation is
predicting the light distribution.  Hence we need a prescription for
where the stars are forming.  Furthermore, it is clear that star
formation is a self-regulating process due to the injection of energy
from supernova, ionizing radiation and stellar winds into the
star-forming gas.  These processes are all happening well below the
resolution scale of even the highest resolution cosmological
simulations, so a sub-grid model is needed to include their effects.
\changa{} includes the
physics of metal lines and molecular hydrogen cooling
 \cite{shen10,christensen12a} and feedback from supernovae (SNe). In
\changa{}, we have implemented the ``blast-wave'' and
``superbubbles'' feedback models described in \cite{Stinson06} and
 \cite{keller14} respectively. In both models SF occurs in high gas
density regions and the time distance scale for energy injection into
the gas is then determined by physically motivated models. The
``blastwave'' prescription follows an analytic model of the Sedov blast
wave and it has allowed us to successfully model a number of trends in
galaxy populations including the Tully-Fisher
relation \cite{governato07}, the mass-metallicity relation
 \cite{Brooks07}, the stellar mass-halo mass relation \cite{Munshi13}
and the formation of DM cores in dwarf galaxies \cite{governato12}.


\section{Parallelization Approach}
\label{sect:parapproach}
In \changa{}, the particle distribution in space is represented in the form of a
hierarchical tree structure where each node represents a portion of the 3D space
containing the particles in that volume. The root node represents the entire
simulation space and the children represent sub-regions. The leaf nodes of the
tree are \emph{buckets} containing a small set of particles.

\subsection{Domain Decomposition}
During domain decomposition, particles are divided among objects called
\emph{tree pieces} (or chares in the context of Charm++) which are mapped onto
processors by the runtime system.  Typically, there are more tree pieces than
the number of processors to benefit from the overlap of communication with
computation and the load balancing features enabled with over-decomposition.

\changa{} supports various domain decomposition techniques, which have been
evaluated previously ~\cite{AmitThesis}. We used space-filling curve (SFC)
decomposition for the results in this paper as that is the method currently used
for most scientific studies with \changa{}.

The goal of this scheme is to identify a set of splitting points
(\emph{splitters}) along the space filling curve such that each range contains
approximately equal number of particles. The algorithm used to identify the
splitter keys is similar to the parallel histogram sort~\cite{ParSort09}. First,
a single master object calculates a set of splitters along the SFC that
partition the simulation domain into disjoint areas of roughly equal volume. It
then broadcasts the splitter keys to all the tree pieces. The tree pieces
evaluate the count of particles for each bin, which is reduced across all tree
pieces back to the master process. The candidate keys are then adjusted based on
the contributions received, and new splitters are broadcast for any bins that
are not sufficiently close to an optimal partition. This process is repeated
until a suitable set of splitter keys is determined such that all tree pieces
have roughly equal number of particles. After the splitter keys are identified,
particles are globally distributed to tree pieces according to the splitters,
where each bin corresponds to one tree piece.


\subsection{Tree Build}
After particles have migrated and domain decomposition is finished,
each tree piece builds its tree independently. The tree build is done
in a top-down manner. The algorithm starts from the root, which contains the entire
simulation space, and proceeds downwards to the leaves, which are buckets
containing a small number of particles, typically 8 to 12. A tree piece has information about the extent
of the domain held by other tree pieces; this information is used in the tree building
process. 
A spatial binary tree is then constructed by bisecting the
bounding box containing particles in the given volume. The tree building
process bisects each node, starting at the root, into children,
which represent sub-regions within the space, until a leaf node is constructed.
If a node in the
tree held by a tree piece contains particles in another tree piece, then that node
becomes a boundary node.

We also take advantage of the fact that a tree piece can access other tree pieces
within the same address space. All the tree pieces within the same
address space are merged.  After the merge, each tree piece has read-only access to the tree
datastructure that is constructed by merging multiple tree pieces. For additional
details, we refer the reader to~\cite{2007_ChaNGaScaling}.

\subsection{Tree Traversal and Gravity}
The object of tree traversal is to identify for each bucket of particles in the
tree the list of nodes and particles whose information is needed for the gravity
calculation. These \emph{interaction lists} are constructed on a per bucket
basis to amortize the overhead of the tree traversal.


Another optimization that is implemented in \changa{} to improve the performance of
the gravity phase is based on the observation that nearby buckets tend to have
similar interaction lists~\cite{Stadel01}. The algorithm constructs the
interaction list of a parent node before proceeding to the children, and
maintains a \emph{checklist}, passed down the tree, that reduces the number
of nodes that need to be evaluated at each level.

\begin{figure*}[t]
\includegraphics[width=.96\textwidth]{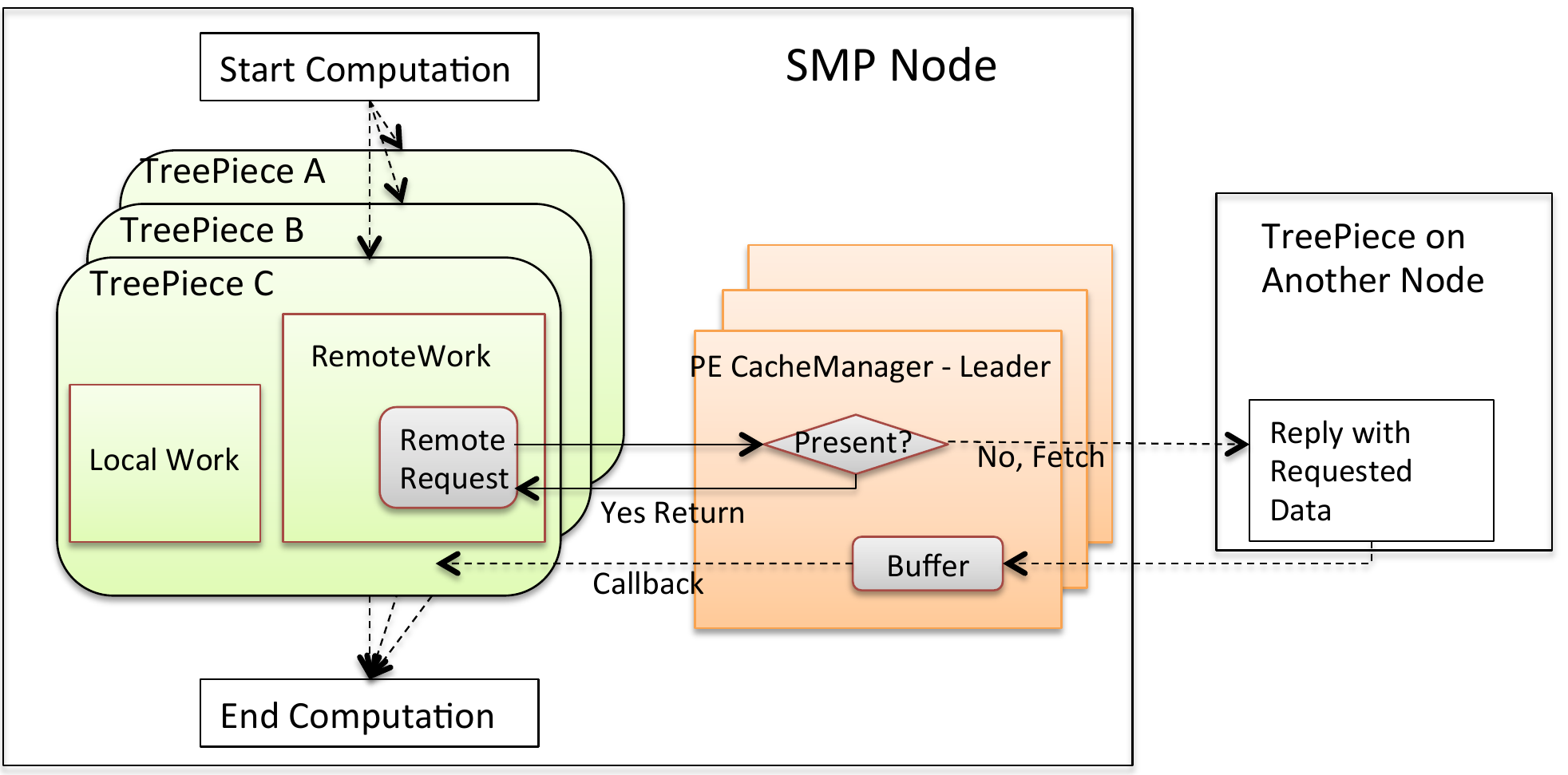}
\caption{An overview of the gravity force calculation in \changa{} with a software cache.}
\label{fig:gravity}
\end{figure*}

Tree traversal results in remote access of nodes which are part of
tree pieces on other processors. To optimize this remote data access, we have
implemented a software cache, as shown in Figure~\ref{fig:gravity}. 
The \emph{Cache Manager} serves node and particle requests
made by a tree piece. If a node request is missed in the cache, then a
request is sent to the corresponding tree piece. If there is already an
outstanding request in the cache, no additional request is sent. When the
response arrives, the requestors are informed and the walk resumes. This
improves the performance by hiding the latency of remote requests by improving
the chance of a requested node being already present in cache as well as
reducing the number of messages sent and received for the remote node. To
further reduce cache misses, we also perform a prefetch walk which obtains remote
node information.

To effectively overlap communication and computation, we divide the tree
traversal into local and remote. A local traversal is done on the portion of the
tree which is within the local address space whereas a remote traversal is done on the remaining
part of the tree which requires communication between the tree pieces. We use
prioritization to give precedence to the remote traversal, which requires
communication, over the computation-dominated local traversal. When the
remote walk has sent out requests for the node and is waiting for the response,
the local walk can be done. This enables overlap of communication with local
computation and helps mask message latency. Figure~\ref{fig:gravity} diagrams
the gravity calculation in \changa{} with a software cache.

Sequential code in \changa{} is also well optimized. In particular, we take
advantage of single-instruction, multiple-data (SIMD) parallelism inherent in
the force calculation to accelerate that part of the computation using FMA or SSE
vector instructions.



\section{Datasets and Systems}
\label{sect:dataset}

We first describe the datasets used for our experiments and their
characteristics.  We have two large, uniform (Poisson distributed)
datasets with 12 and 24 billion particles.  Other than having periodic
boundary conditions these two datasets are not particularly
interesting for cosmology.  We include them here to demonstrate the
scaling of \changa{} to large core counts.  \emph{cosmo25} is the more
challenging dataset: it is a 2 billion particle snapshot taken from
the end (i.e. representing the current, very clustered, structure of
the Universe) of a
dark matter simulation of a 25 Megaparsec cube in a $\Lambda$CDM
Universe.  The force softening is 340 parsecs, and the simulation
represents a challenge for load balancing.  The
version of this
simulation with gas dynamics and star formation is able to resolve the
disks of spiral galaxies within this volume [Anderson et al, in preparation].
\emph{dwarf} is our most challenging dataset: while it contains only
52 million particles spread throughout a 28.5 megaparsec volume, most
of the particles are in a single high resolution region in which a
dwarf galaxy is forming.  The mass resolution in this region is
equivalent to having 230 billion particles in the entire volume, and
the force resolution within this region is 52 parsecs.  This is a high
resolution version of the DWF1 simulation discussed in \cite{governato07}.
\\

We show the performance of \changa{} on Blue Waters.
Blue Waters is a hybrid Cray XE/XK system located at the National Center for Supercomputing
Applications (NCSA). It contains $22,640$ Cray XE6 nodes and $4,224$ Cray XK7 nodes
that include NVIDIA GPUs. Each dual-socket XE6 compute nodes contains two AMD Interlagos
$6276$ processors with a clock speed of $2.3$ GHz  and $64$ GB of RAM. 


\section{Single Stepping}
\label{sect:singlestep}
\begin{table}[t]
\newcolumntype{C}{>{\centering\arraybackslash}X}
\begin{tabularx}{\columnwidth}{|C|C|C|C|C|C|}
\hline
{\bf \#cores} & {\bf Gravity} & {\bf DD} & {\bf TB} & {\bf LB} & {\bf Total Time} \\ 
\hline
16384 & 77.556 &  1.299 & 0.729 & 0.128 & 79.712 \\ 
32768 & 39.254 &  0.698 & 0.617 & 0.136 & 40.705 \\
65536 & 19.876 &  0.496 & 0.367 & 0.062 & 20.801 \\
131072 &  9.967 & 0.181 & 0.138 & 0.027 & 10.313 \\ 
262144 &  5.051 & 0.109 & 0.076 & 0.013 & 5.249 \\ 
524288 &  2.569 & 0.073 & 0.034 & 0.008 & 2.684 \\ 
\hline
\hline
32768 & 75.090 & 1.553 & 0.735 & 0.186 & 77.564 \\ 
65536 & 37.941 &  0.787 & 0.462 & 0.111 & 39.301  \\
131072 &  19.062 &  0.428 & 0.245 & 0.063 & 19.798  \\
262144 &  9.682 & 0.232 & 0.152 & 0.042 & 10.108  \\
524288 &  4.903 & 0.146 & 0.095 & 0.022 & 5.166 \\
\hline
\end{tabularx}
\caption{Breakdown of time for 1 step in seconds for $12$ billion particles (top
half) and $24$ billion particles (bottom half) datasets run on Blue Waters with the proposed optimizations.}
\label{tab:bw12btime}
\end{table}

\begin{table}[t]
\newcolumntype{C}{>{\centering\arraybackslash}X}
\begin{tabularx}{\columnwidth}{|C|C|C|C|C|C|}
\hline
{\bf \#cores} & {\bf Gravity} & {\bf DD} & {\bf TB} & {\bf LB} & {\bf Total Time} \\ 
\hline
16384 & 82.424 & 2.81 & 0.995 &7.79  &94.019 \\
32768 & 42.712 & 1.966 &1.005 &6.854 &52.537 \\
65536 & 21.438 & 1.731 &0.729 &6.482 &30.38 \\
131072 & 12.162 & 1.674 &0.803 &5.718 &20.357 \\
\hline
\hline
32768 & 80.144 &  2.859 & 1.366 & 16.173 &  100.542 \\
65536 & 41.279  & 2.356 & 1.032 & 9.338 & 54.005 \\
131072 &  22.958  & 2.142 & 1.018 & 8.854 & 34.972 \\
\hline
\end{tabularx}
\caption{Breakdown of time for 1 step in seconds for 12 billion particles (top
half) and $24$ billion particles (bottom half) dataset run on blue waters
without the proposed optimizations.}
\label{tab:bw12b24bnoopt}
\end{table}

We now describe essential optimizations required for scaling the simpler
datasets that are not highly clustered, and evaluate their performance. Later
sections will describe optimizations for clustered datasets.

\subsection{Single Stepping Improvements}
We observed that from-scratch domain decomposition is not required at every step,
especially for datasets which are not highly clustered. 
After the initial domain decomposition, it needs to be
performed only when there is an imbalance in the load of tree pieces. By reusing
the previously determined splitters, we reduce the overhead incurred in finding
the splitters as well as the number of particle migrations.  We use an
adaptive mechanism to determine when to perform the domain decomposition. In this approach,
load statistics of the tree pieces are collected and domain decomposition is only
performed if an imbalance is detected. Otherwise, only particle migration is
done based on the previous splitters. We use the quiescence detection~\cite{QuiescenceINTL94} mechanism
implemented in Charm++ to determine when all the migrations are finished. 

In the unoptimized version of the code, the tree build requires all tree pieces to send
the information about the  first and the last particle in their domain, subject to the SFC. This
information is used to determine ownership of nodes in the tree but requires
heavy communication. We avoid this by using the boundary information to
determine a set of candidate tree pieces which may have information about the
node. One of them is then queried and in case that tree piece does not have the
information, it forwards it to the appropriate tree piece. 

Since load balancing incurs overhead, it should be done sparingly. We use the
\emph{MetaBalancer}~\cite{MetaBalancer12} framework in Charm++ to determine when
to invoke the load balancer. \emph{MetaBalancer} monitors the application
characteristics and invokes the load balancer only when needed. 

\subsection{Performance}

\begin{figure}[t]
\includegraphics[width=.96\textwidth]{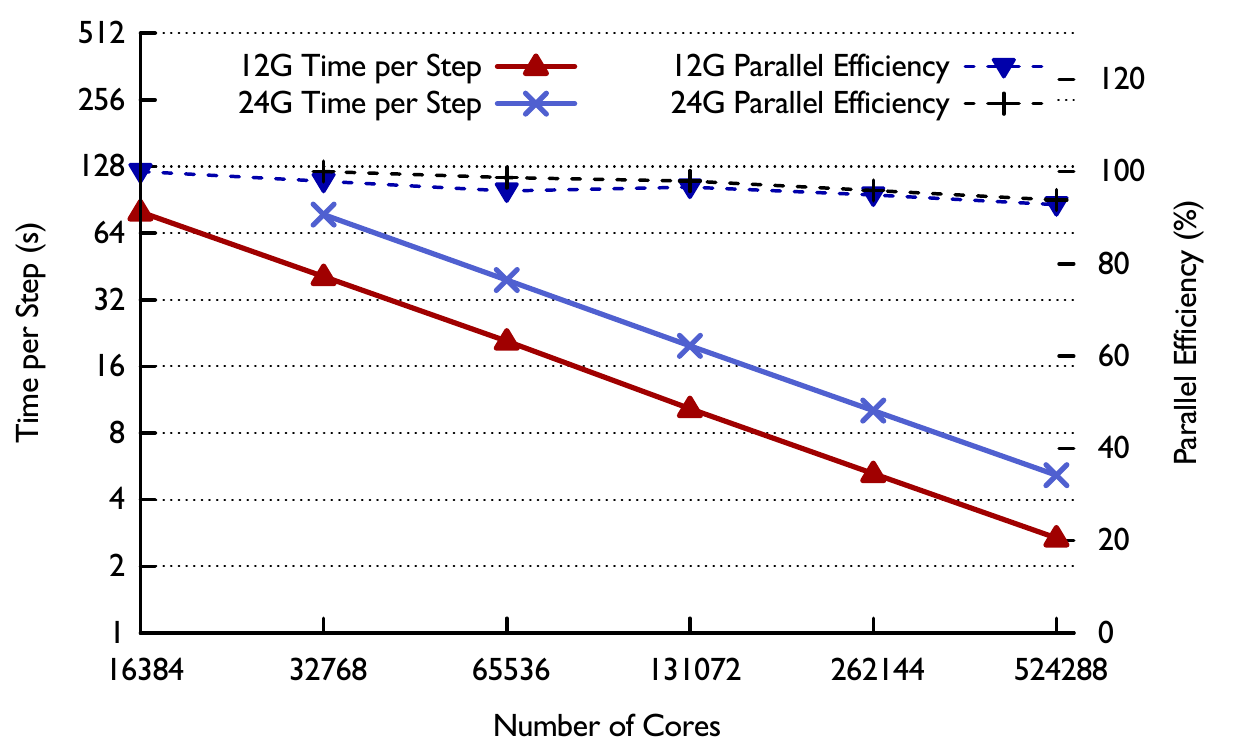}
\caption{Time per step and parallel efficiency for $12$ and $24$ billion
particles on Blue Waters. Both the cases scale well achieving a parallel
efficiency of $93\%$.}
\label{fig:bw12a24bscaling}
\end{figure}

\begin{figure}[t]
\includegraphics[width=.96\textwidth]{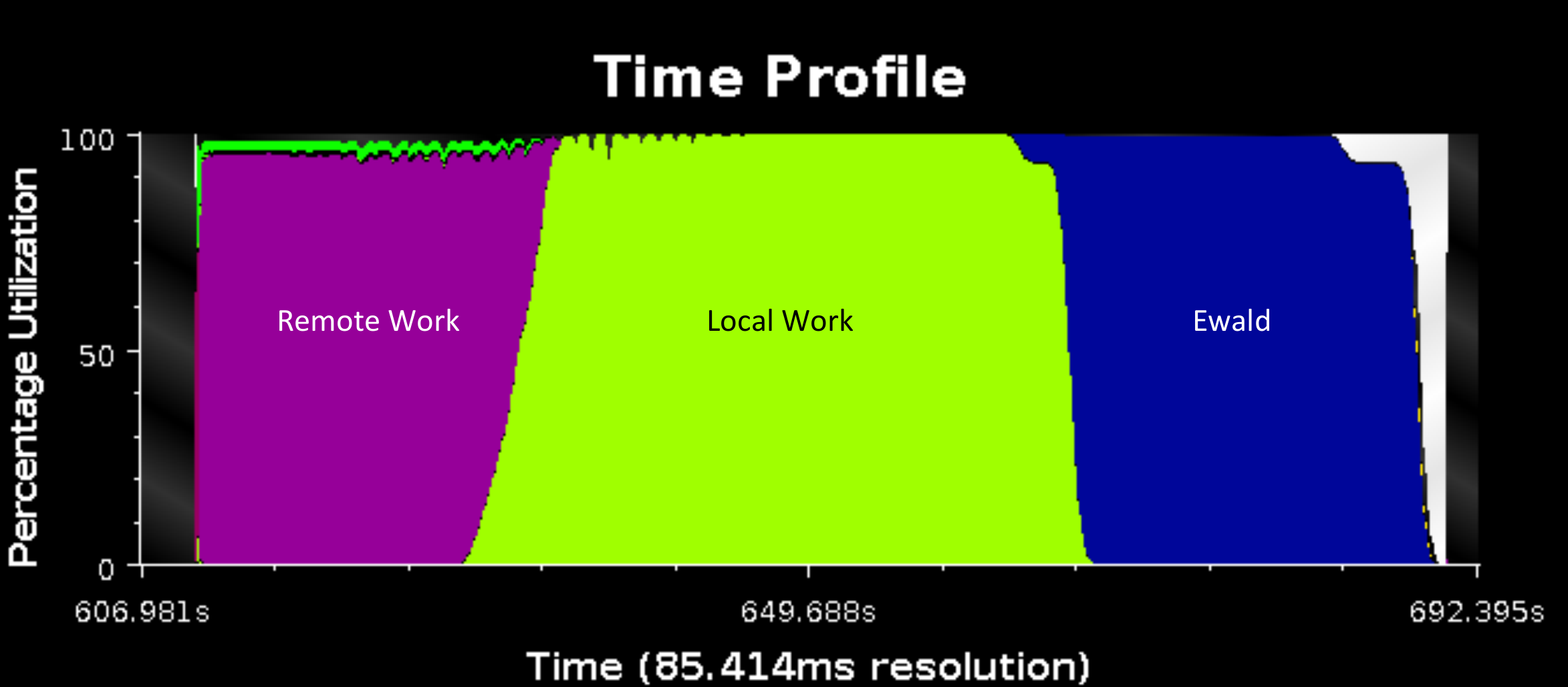}
\caption{Time profile graph which shows processor utilization over time for
$16K$ cores on Blue Waters for $12$ billion particles. This shows overlap of
communication with computation to achieve high utilization.}
\label{fig:bw12timeprofile}
\end{figure}

Figure~\ref{fig:bw12a24bscaling} shows
strong scaling results on up to $512K$ cores on Blue Waters evolving $12$ and $24$ billion
particles. Our application exhibits almost perfect scaling up to the maximum number of cores. 
Each iteration consists of domain
decomposition, load balancing, tree build and force calculation. Table~\ref{tab:bw12btime} shows
the break down of the time per step into the different phases. 
For the simulation evolving $12$ billion particles, we achieve $93\%$ parallel efficiency
at $512K$ cores with the time per step being $2.6$ seconds. 
For the $24$ billion particles, we achieve $93.8\%$ parallel efficiency with a
time per step of $5.1$ seconds. The
efficiency is calculated with respect to $16K$ cores and $32K$ cores for $12$
and $24$ billion respectively. 

The good scaling of the gravity phase is due to the overlap of communication and
computation, the improved tree walk algorithm using an interaction list, the
software request cache, prefetching, and other optimizations. The time for
domain decomposition also scales with the increase in number of cores.
Table~\ref{tab:bw12btime} shows, for the $12$ billion particles at $512K$ cores
on Blue Waters, that domain decomposition takes on average $73$ ms per step. At
$128K$ cores the domain decomposition is $9$ times faster in comparison to the
unoptimized version. This is due to the use of the adaptive technique to
determine when to perform full domain decomposition. The tree build time also
scales well and takes $34$ ms at $512K$ cores. At $128K$ cores, the tree build
is approximately $6$ times faster than the unoptimized version.  Similar trends
are seen in the $24$ billion particle simulation.

Table~\ref{tab:bw12b24bnoopt} contains the breakdown of the
total time per step for the unoptimized version of the code. Comparing the
results with table~\ref{tab:bw12btime}, for the $12$
billion particle simulation, we reduce the total time by
$15$ to $49\%$.  For the $24$ billion particle simulation, we
reduce the total time per step by $22$ to $43\%$.
The reduction in time occurs for all phases of the application.

Figure~\ref{fig:bw12timeprofile} shows the time profile graph obtained using
Projections~\cite{Projections}. This shows the average processor utilization
over the course of one time step evolving $12$ billion particles on $16K$ cores of
Blue Waters. We can see that the local work, which is given a lower priority,
overlaps with the communication needed for the higher-priority remote work, resulting
in close to $100\%$ processor utilization.

\section{Clustered Dataset Challenges}
\label{sect:dwrf}
\begin{figure}[t]
\centering
\begin{tabular}{@{}c@{}c@{}}

  \subfloat[Without replication] {
  \includegraphics[width=0.47\textwidth, trim=0 0 0 10]{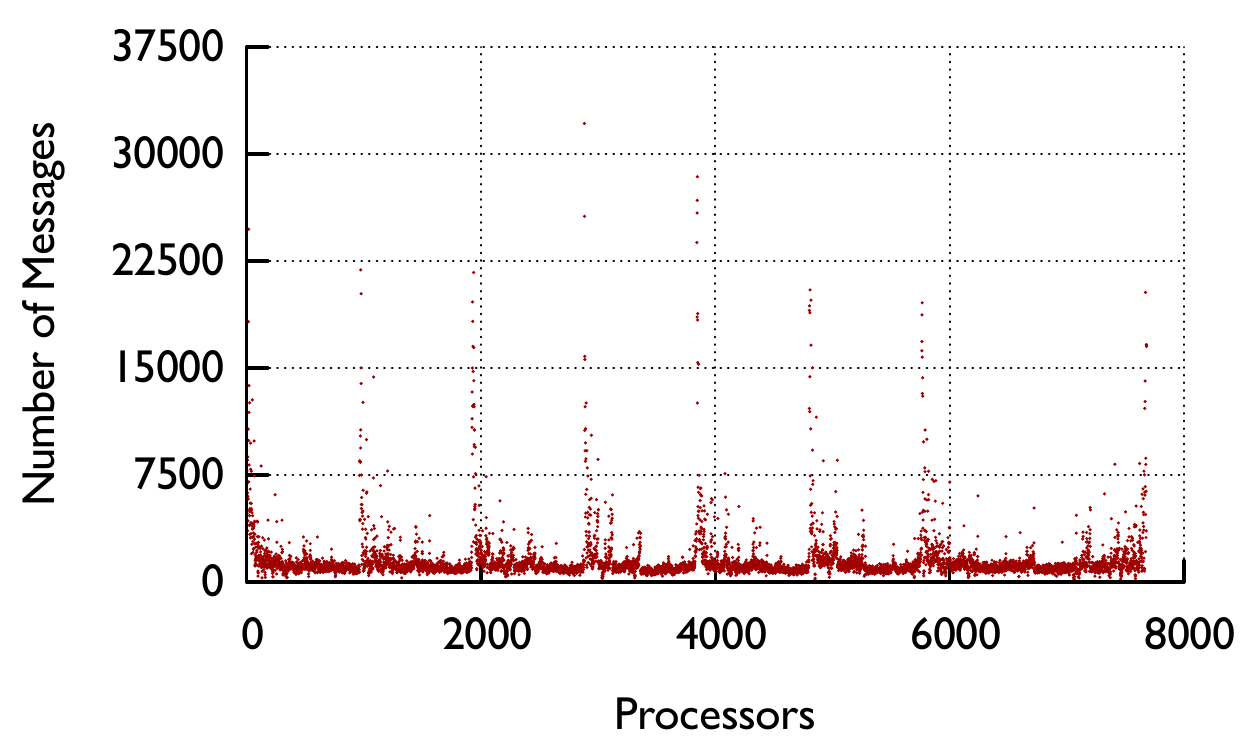}
  \label{fig:dwfmsgworepl}
  } &
  \subfloat[With replication] {
  \includegraphics[width=0.47\textwidth, trim=0 0 0 10]{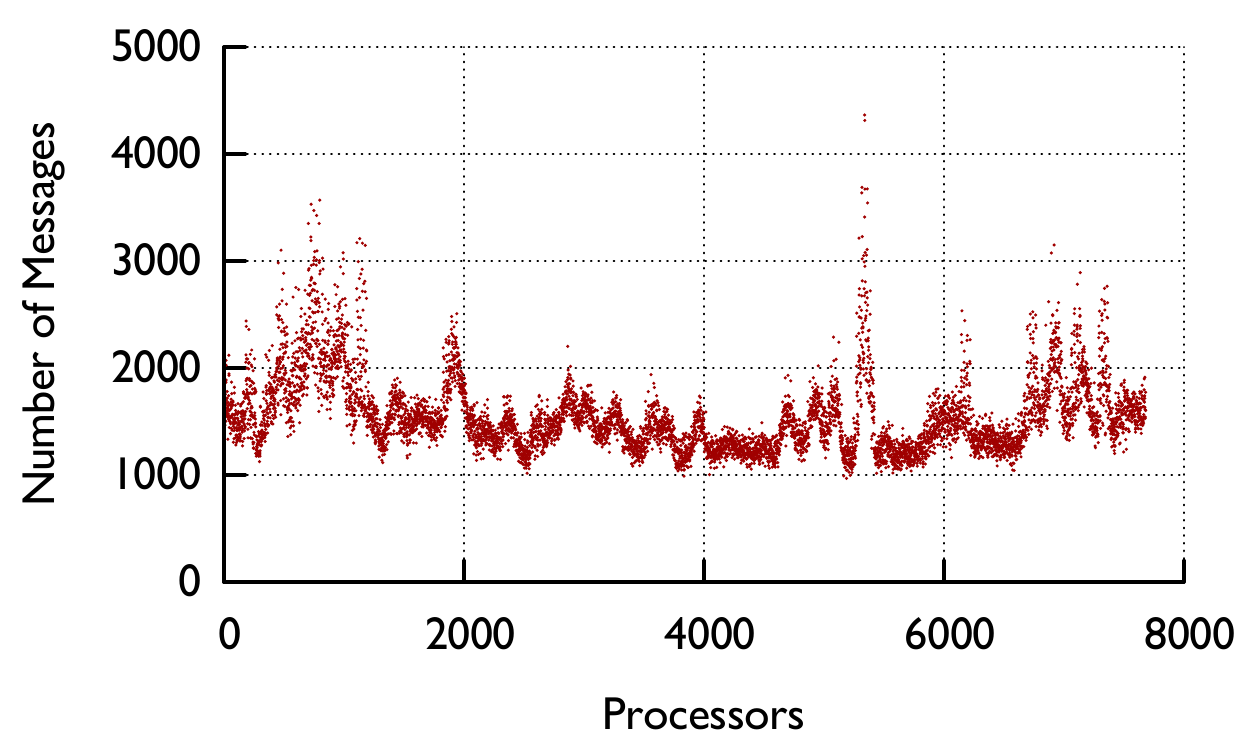}
  \label{fig:dwfmsgw4repl}
  } 
\end{tabular}
\caption{Number of messages received by processors for a simulation of
\emph{dwarf} dataset on $8K$ cores on Blue Waters. Note that replication
reduces the maximum request received by a processor from $30K$ to $4.5K$.}
\label{fig:dwfcommmsg}
\end{figure}

\begin{figure*}
\centering
\begin{tabular}{@{}c@{}c@{}}

  \subfloat[Without replication] {
  \includegraphics[width=0.47\textwidth, trim=0 0 0 10]{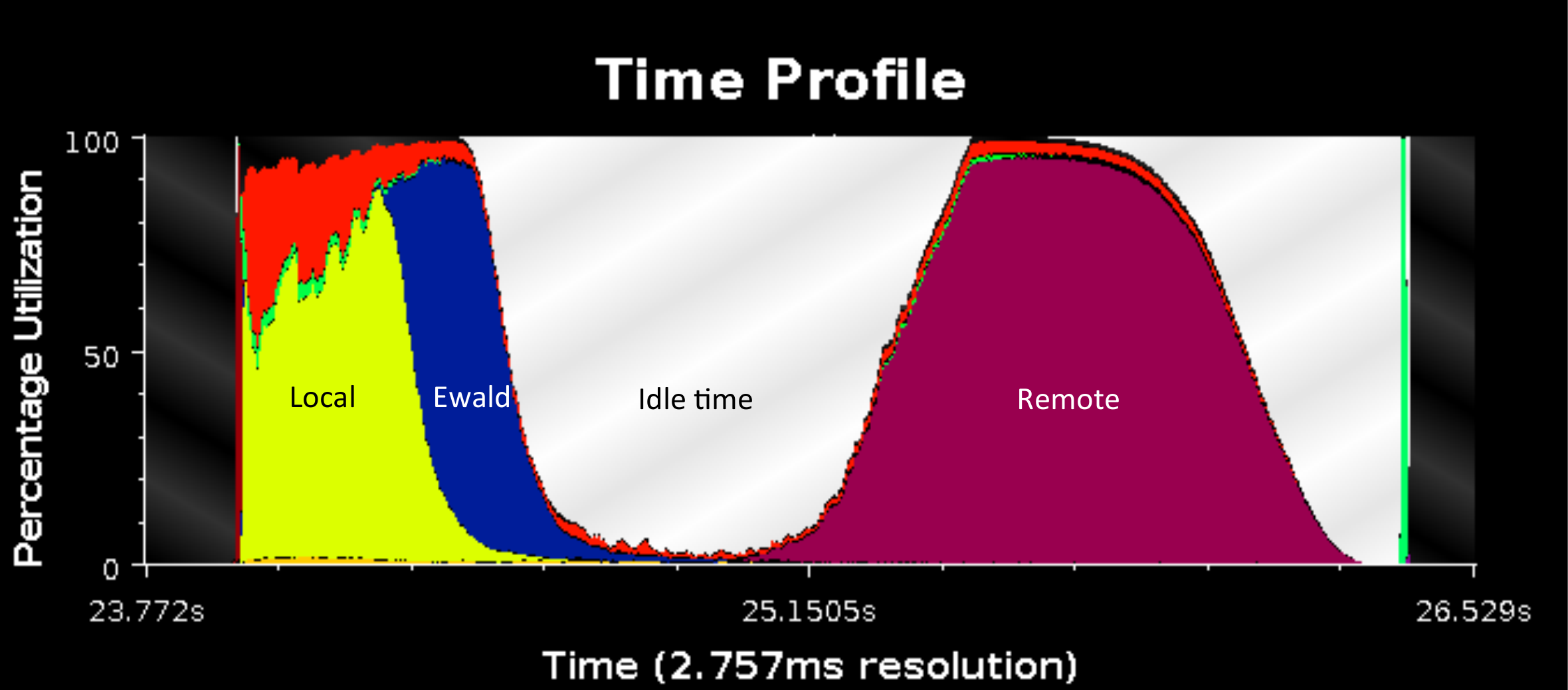}
  \label{fig:dwfcommworepl}
  } &
  \subfloat[With replication] {
  \includegraphics[width=0.47\textwidth, trim=0 0 0 10]{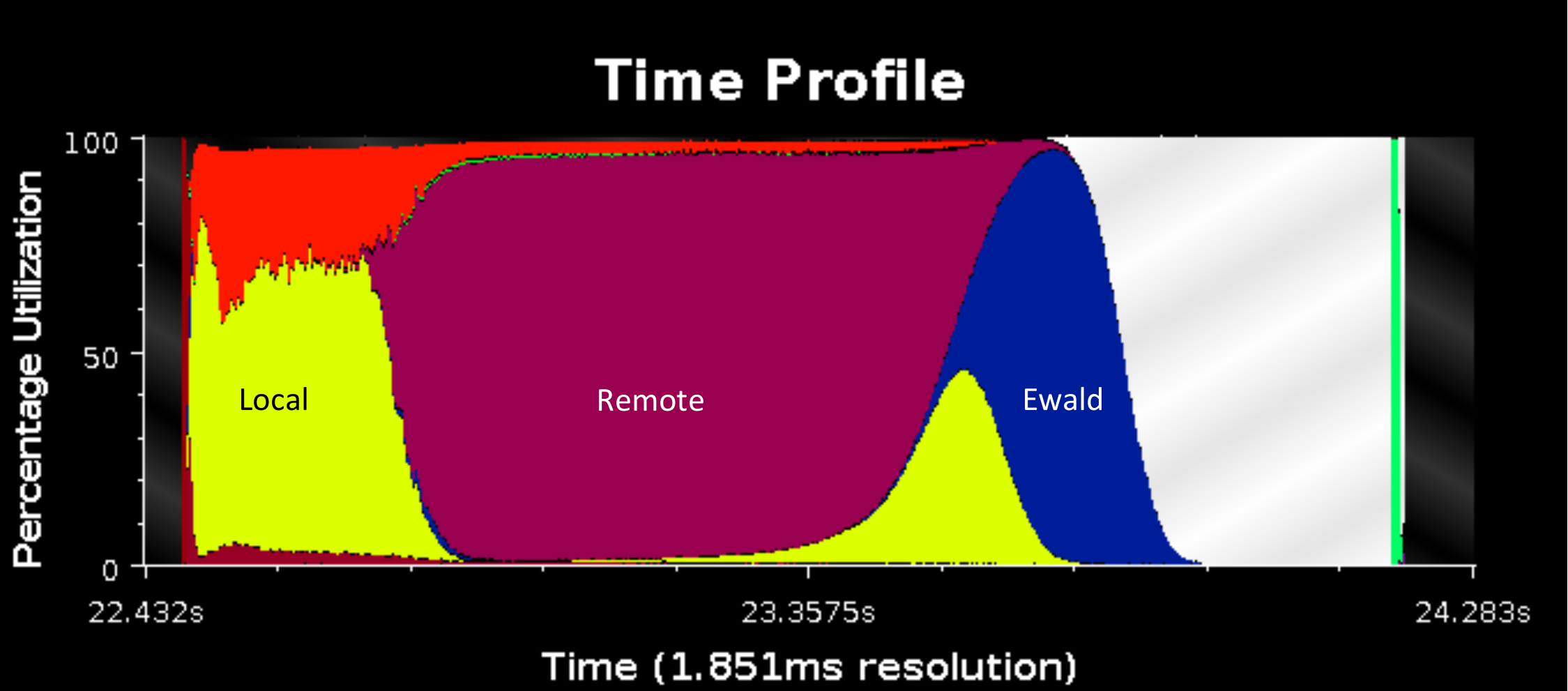}
  \label{fig:dwfcommw4repl}
  } 
\end{tabular}
\caption{Time profile graph showing processor utilization over time for
simulation of \emph{dwarf} dataset on $8K$ cores. Note the idle time without
replication which is removed by the replication and the gravity time is improved
from 2.4 seconds to 1.7 seconds.}
\label{fig:dwfcomm}
\end{figure*}

\begin{figure}[t]
\includegraphics[width=.96\textwidth]{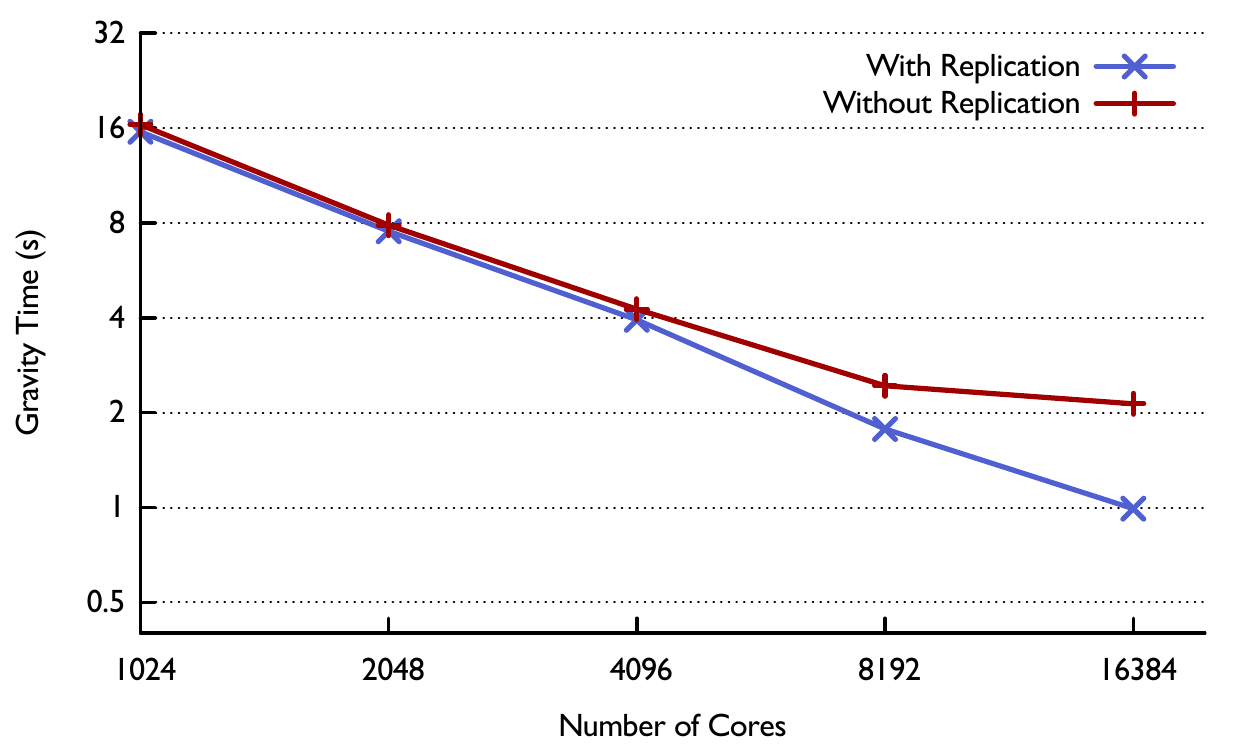}
\caption{Gravity time for the \emph{dwarf} simulation on $8K$ cores on Blue
Waters with and without the replication optimization.}
\label{fig:dwfscaling}
\end{figure}
For datasets such as \emph{dwarf}, the particle distribution is concentrated at
the center and therefore highly clustered.
This creates many challenges in scaling, of which one of the most significant 
is communication imbalance. During
the gravity phase, remote requests are sent for tree nodes that are not present in
the local cache. In a clustered dataset, some tree nodes are requested many more
times than others. This results in the tree pieces owning those tree nodes
receiving a large volume of node request messages.  
Figure~\ref{fig:dwfmsgworepl}
shows the number of requests received by processors for the \emph{dwarf}
simulation at $8K$ cores on Blue
Waters. We can see that a handful of processors receive as many as $30K$ messages.
Even though there is overlap of communication with
computation,
this causes significant performance degradation. This is because, at this scale,
there is not enough local computation to overlap seconds of delay in
receiving messages. One way to mitigate this problem is to replicate the
information that is being requested to prevent a few processors from being the
bottleneck. 

We replicate the information about the tree nodes on multiple processors
ensuring that no single processor becomes overloaded.
Before the gravity phase begins,
tree pieces send their node information to a set of \tprs{} on other
processors. The responsibility of the \tpr{} is to store the node information
sent to it and handle requests for
those nodes. When a tree piece needs to request for a remote node, it chooses
randomly one of the \tprs{} to send the request to. 
Figure~\ref{fig:dwfmsgw4repl} shows the number of messages received by the
processors when four \tprs{} are created for each tree piece. 
For $8K$ core run on Blue Waters, replication reduces the maximum number of
messages received from $32K$ to $4.2K$ and the requests are better distributed
among all processors.  Figure~\ref{fig:dwfcomm} shows the time-profile graph
where the x-axis is the time and y-axis is the processor utilization.
Here, yellow regions constitute the local work, blue the ewald and maroon the
remote work. 
Note the idle time, in figure~\ref{fig:dwfcommworepl}, before the remote work
begins which is due to the delay in receiving messages and with no local work
overlap. Figure~\ref{fig:dwfcommw4repl} shows the impact of replication. The
remote work can start earlier due to a smaller delay in request messages.
The local work overlaps with the communication until remote work is ready to start.
This is a very
good example to show prioritization of remote work over local work and the overlap
of communication with computation.
Figure~\ref{fig:dwfscaling} shows the strong scaling performance for this
dataset on core counts ranging from $1K$ to $16K$. We compare the time for the
gravity phase because the rest of the phases are the same in both cases.  The gravity time is
improved  from $2.4$ seconds to $1.7$ seconds for $8K$ cores and from $2.1$ seconds to $0.99$
seconds on $16K$ cores. 

%
%
%

\section{Multi-stepping Challenges}
\label{sect:multistep}

A wide variation in mass densities can result in particles having dynamical
times that vary by a large factor. In a single-stepping mode, good accuracy can
only be achieved by performing the force calculation and particle position and
velocity updates at the smallest timescale. However, hierarchical time stepping
schemes can be used for a large dynamic range in densities at a low additional
cost. We use adaptive time scales where forces are evaluated only on relevant
particles instead of evaluating forces on all the particles at the smallest time
scale. In a multi-step simulation, particles are assigned to time step rungs
corresponding to the shortest time scale required for accurate simulation. Rungs
corresponding to short time scales are evaluated more frequently than those for
long time scales.

Using multi-stepping for clustered datasets introduces a variety of
challenges. The irregular distribution of particles in the simulation space as
well as the division of particles into rungs creates severe load imbalance. In
general, the challenge is higher for datasets with fewer particles. We discuss
various optimizations that enable \changa{} to scale a medium-sized $2$ billion
particle clustered dataset, \emph{cosmo25}, on up to $128K$ cores on Blue
Waters. Reaching this level of performance required overcoming
challenges related to load imbalance, communication overhead with a decrease in
computation per processor as well as the scalability of other phases of the
simulation. Strong scaling of this nature will be required to run clustered
cosmological simulations on future machines with hundreds of Petaflop/s
performance, and presents a realistic proving ground for parallel strategy
innovations.

\subsection{Optimizations for the Gravity Phase} 
\label{sect:optgrav}

\begin{figure}
\centering
\begin{tabular}{@{}c@{}c@{}}

  \subfloat[Rung 0] {
  \includegraphics[width=0.47\textwidth, trim=0 0 0 10]{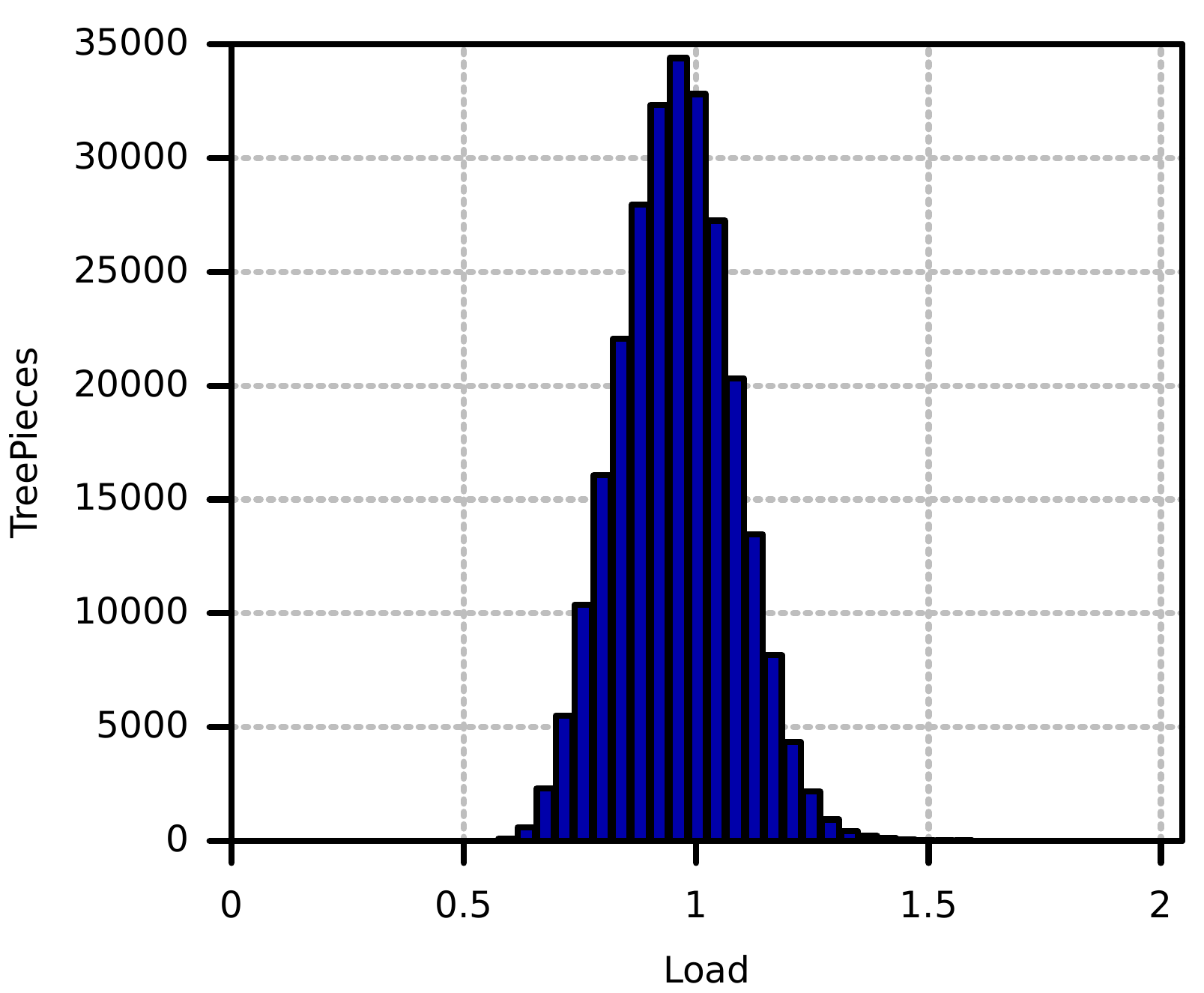}
  \label{fig:rung0dist}
  } &
  \subfloat[Rung 4] {
  \includegraphics[width=0.47\textwidth, trim=0 0 0 10]{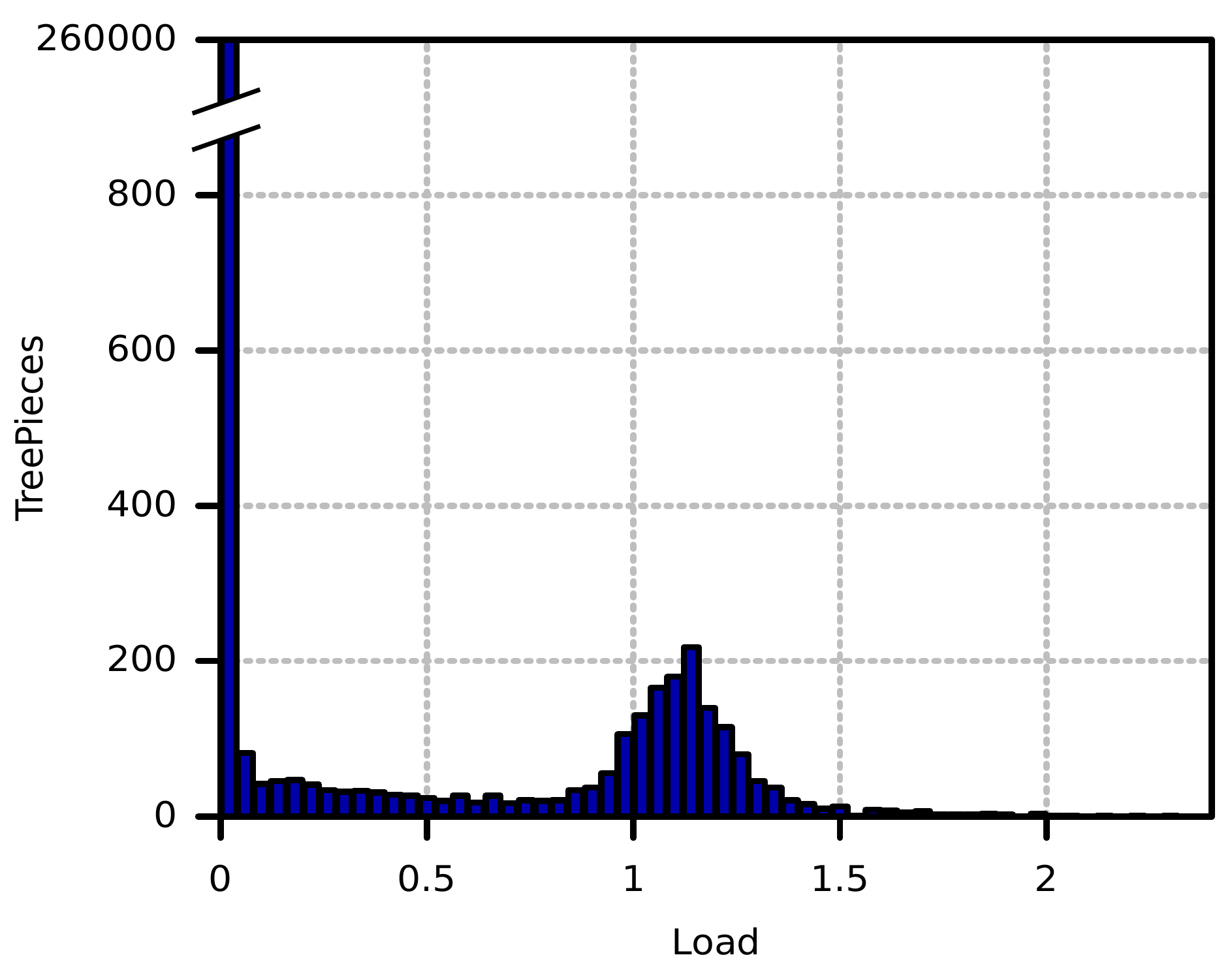}
  \label{fig:rung4dist}
  }  
\end{tabular}
\caption{Distribution of tree piece load for rung 0 (slowest) and rung 4
(fastest). Rung 0 has load distributed around the mean. Rung 4 has only $2405$
active tree pieces with a maximum load of $2.3$}
\label{fig:rungdist}
\end{figure}

In a multiple time step simulation, the number of particles active in the fastest rung is
typically only a fraction of the total number of particles being simulated.
These active particles tend to be clustered, and therefore the distribution of
particles among the tree pieces is highly imbalanced. One may consider performing
from-scratch domain decomposition based on the active set of particles for these
time steps but that results in large jumps of the domain boundaries. To prevent such sudden
large variations of the boundaries, we perform from-scratch domain decomposition
only when there is a significant number of particles active for that time step. But
as one can imagine, this will result in tree pieces with a large variation in
active particles and load. Figure~\ref{fig:rungdist} shows the
distribution of the 
load on tree pieces for the fastest rung (rung 4) and the slowest rung (rung 0) of
the \emph{cosmo25} dataset.  The slowest rung has tree pieces with loads
distributed around the mean.  But the fastest rung has only $2405$ tree pieces with
active particles and some of them have a load which is $3000$ times the average
load of tree pieces and $40$ times the average load of the system.  Even though
periodic load balancing is performed to distribute the load, the maximum load of
the system will be limited by the most overloaded processor which in this case
is the one having the most loaded tree piece. At larger scales of $128K$ cores
there is not enough work to be distributed among all the cores which results in
significant degradation of performance. We propose two adaptive strategies to
overcome this problem.
\\
%

\noindent\textbf{Intra-node Work Pushing}\\

We use the SMP mode of Charm++ to take advantage of the shared memory
multiprocessor nodes used in HPC systems~\cite{CharmSMPOptimization}.  The SMP
mode supports multi-threading, where one Charm++ process is assigned per SMP
node, with a single thread mapped to each physical core. One thread within a
node is normally assigned as a communication thread responsible for internode
communication, while the rest are used as worker threads that
implement \emph{processing elements (PEs)}.

Within a
Charm++ SMP process, data can be shared via pointers.  The load balancing strategy
works in a hierarchical fashion. Details are given in Section~\ref{sect:hlb} but
in essence it first tries to achieve load balance among the SMP processes and then
balances the load among cores within the SMP process.


\lbmgr{}, which is an object present on each \pe{}, has information about the average
load of the system and load of other \pes{} on the same SMP process. The \lbmgr{}, on
identifying that a \pe{} is overloaded, instructs overloaded tree pieces at that \pe{} to
distribute the work among other less loaded \pes{} within the SMP process. A tree piece
is responsible for calculating forces on a set of particles in its domain, grouped into
buckets. We consider the bucket to be the smallest entity of work
that can be distributed. \pes{} receiving a foreign bucket
have access to the tree and all the data structures of the owner
tree piece so that they can perform the tree traversal and gravity force calculations
for the foreign bucket. Once the force calculations are done, the
foreign bucket is marked as complete and the original \pe{} is informed. Once all
the foreign and local buckets are completed, the tree piece is done with
the gravity calculations.

\begin{figure*}[t]
\centering
\begin{tabular}{@{}c@{}c@{}}
  \subfloat[Without work pushing] {
  \includegraphics[width=0.47\textwidth, trim=0 0 0 10]{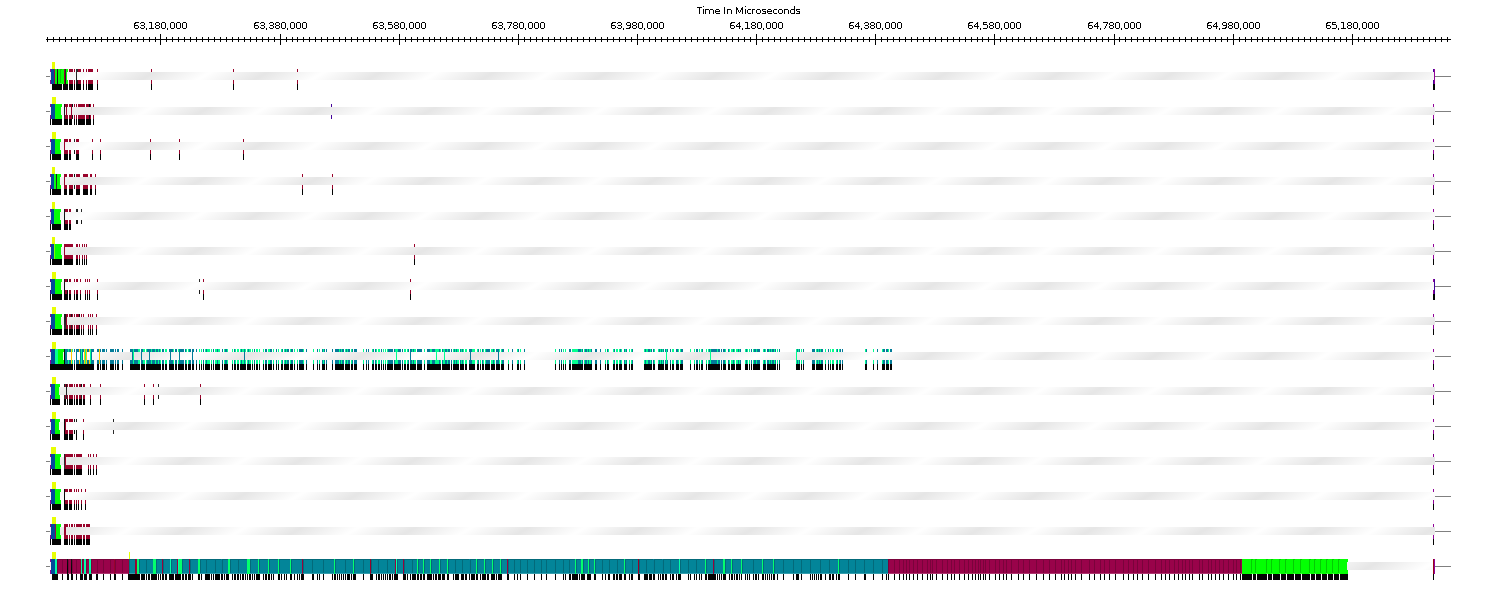}
  \label{fig:woforbuck}
  } &
  \subfloat[With work pushing] {
  \includegraphics[width=0.47\textwidth, trim=0 0 0 10]{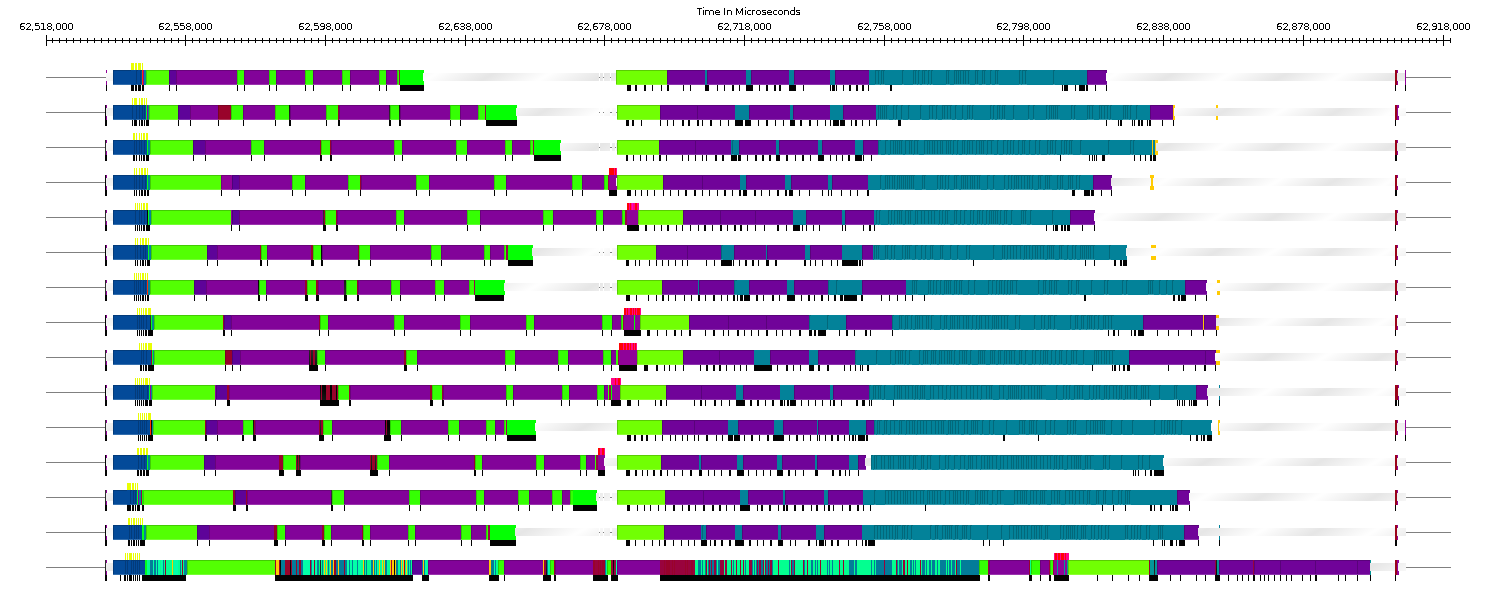}
  \label{fig:withforbuck}
  } 
\end{tabular}
\caption{Time line profile for all the PEs (rows) on a SMP process for the $16K$
cores run. White shows idle time and colored bars indicate busy time. Work
pushing achieves better distribution of work among PEs. The total time per step
reduces from $2.3$ seconds to $0.3$ seconds.}
\label{fig:workpush}
\end{figure*}

This work pushing adaptive strategy reaps the most benefit for time steps
where the fastest rung is active. Figure~\ref{fig:woforbuck} shows the time-line
view from the projections tool~\cite{Projections} for rung 4 (the fastest rung).
Here, each line corresponds to a \pe{} and colored bars indicate busy time while
white shows idle time. This plot is for a $32K$ run on Blue Waters
and we have chosen the \pe{} and the corresponding SMP process with the maximum load.  
We can see that the most loaded \pe{}, which also contains the most loaded
tree piece, is busy for about $2$ seconds while other \pes{} are idle.
Figure~\ref{fig:withforbuck} shows the time line for the work pushing strategy
for a set of \pes{} in the SMP process where one of the \pe{} is assigned the
most loaded tree piece.  With the work pushing strategy, we are able to
successfully distribute the work load among other \pes{} within the node.
This results in a reduction of the gravity time from $2.3$ seconds to $0.3$
seconds for the fastest rung.
\\
\\

\begin{figure*}[t]
\centering
\begin{tabular}{@{}c@{}c@{}}

  \subfloat[Without dynamic rebalancing] {
  \includegraphics[width=0.47\textwidth, trim=0 0 0 10]{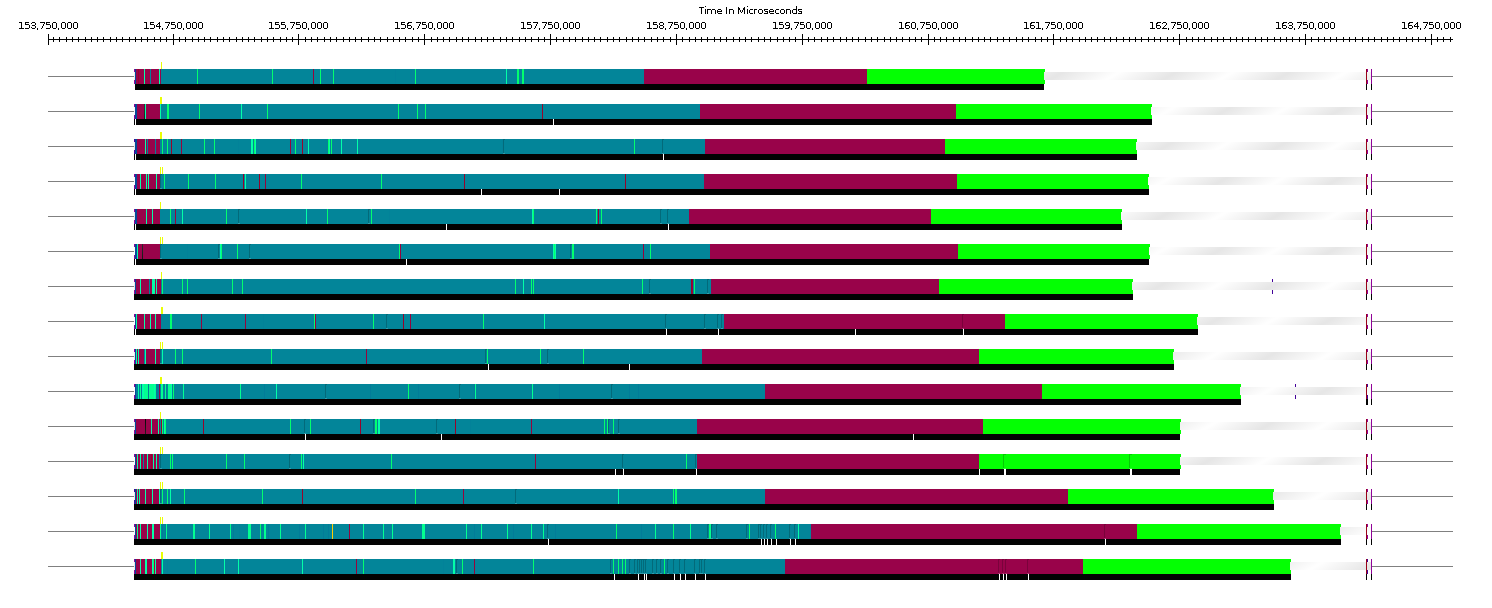}
  \label{fig:wockl}
  } &
  \subfloat[With dynamic rebalancing] {
  \includegraphics[width=0.47\textwidth, trim=0 0 0 10]{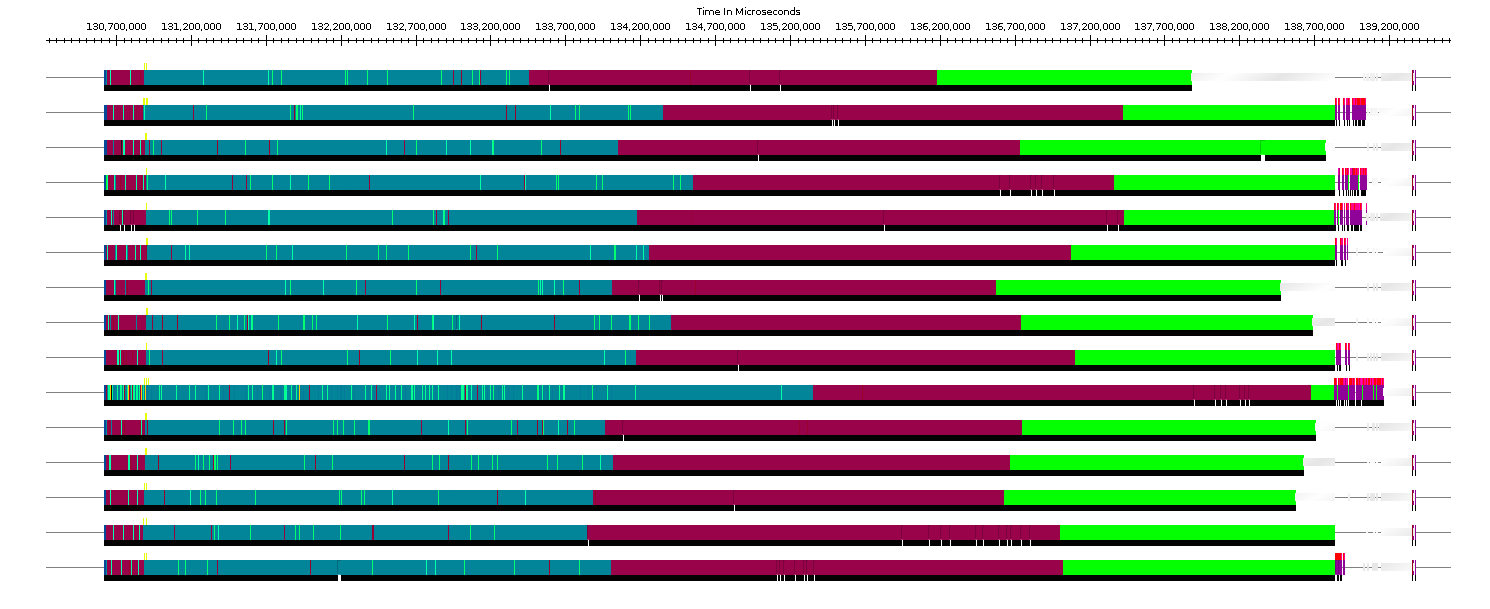}
  \label{fig:wckl}
  } 
\end{tabular}
\caption{Time line profile for all the PEs (rows) on a SMP process for the $16K$
cores run. White shows idle time and colored bars indicate busy time. Dynamic
rebalancing eliminates trailing idle time resulting in better utilization. The
total time per step reduces from $9.8$ seconds to $8.5$ seconds.}
\label{fig:worksteal}
\end{figure*}

%

\noindent\textbf{Intra-node Dynamic Rebalancing}\\
For clustered datasets, it is often the case at the trailing end of the gravity 
calculation that some of the PEs are
idle while others are busy. This could be due to misprediction of load or
inability of the load balancer to balance the load perfectly.
Figure~\ref{fig:wockl} shows the Projections time-line view for this scenario where the colored bars indicate busy
work while the white shows idle time. We found that such slight load
imbalance in the application can be mitigated by more fine-grained parallelism
within the SMP process. We use an
intra-node dynamic rebalancing scheme where the idle PEs within the node pick work
from the busy ones. The scheme is implemented using the \emph{CkLoop} library~\cite{CharmSMPOptimization}
in Charm++, which enables fine-grained parallelism within a SMP process.

As with the work-pushing scheme, buckets are the smallest entity of work that can be reassigned.

If all the tree pieces
residing on a PE have finished their work, then the PE becomes idle.  At each
PE, the \lbmgr{} maintains a PE-private variable which keeps track of its
status.  Since the memory address is shared among the PEs on a SMP process, the
\lbmgr{} can access the status variable of all the PEs within the SMP process.
Whenever there is a significant number of idle PEs, the dynamic rebalancing scheme kicks
in. Tree pieces then create chunks of work out of the unfinished buckets and add these
to the node-level queue.  The idle processors access the node-level queue and
pick up work to execute.  Due to the overhead associated with the node-level
queue we only use the work-stealing scheme adaptively for the
trailing end of the computation.

Figure~\ref{fig:wockl} shows the time-line for the slowest rung, rung 0, of
\emph{cosmo25} dataset simulation for a $32K$ run on Blue Waters. We pick a
subset of PEs to show this problem. We can see that the load is almost balanced but towards
the end of the step there are some PEs which are idle while others are busy.
Figure~\ref{fig:wckl} shows the time-line with dynamic rebalancing. It is able
to successfully handle small amounts of load imbalance and reduce the gravity
time from $9.8$ seconds to $8.5$ seconds for rung 0.


\subsection{SMP Request Cache}
Data reuse can be critical in determining the performance of tree-based
algorithms~\cite{cosmo2006}.  Modern SMP-based supercomputers
offer several levels at which data sharing can be effective.  Requests for the
same remote elements from two traversals on a core can be merged.
The fetched data can then be reused by all traversals on the core.
Similarly, cores in the same SMP domain can share remotely fetched data.
In the following we describe a two-level caching scheme that
enables the data reuse across traversals on a core, as well as across
cores on an SMP processor. This caching mechanism is transparent to the
traversal code.

Each core on the SMP has a {\em private} cache,
which stores pointers to remotely fetched data.
There also exists one cache at the level of the
SMP that is {\em shared} by all cores in the SMP. The
shared cache contains the union of all the entries in the private caches of 
these PEs.

Briefly, the algorithm funnels all requests for remote data through the 
cache. If the data are found in the private cache, then they are immediately
passed into the requesting traversal's visitor code. If the data are not 
found on the PE, we check whether some other piece on the PE has requested
them previously. If so, a lightweight continuation is created to resume the 
traversal at the requested node upon its receipt. Otherwise, the more 
expensive, SMP-wide table lookup is performed. 

We devised a scheme to manage concurrent accesses of the shared,
SMP-wide cache table, where all requests for remote data
generated by traversals on the SMP processor are funneled through a single core, which is termed the
{\em fetcher} for that SMP processor. Cheap, intra-node messaging between PEs is used
for efficiency. 

\subsection{Domain Decomposition}
Simulations of datasets with nonuniform distributions are characterized by
extensive movement of particles across tree piece boundaries over time. When
unchecked, this leads to an increasingly nonuniform distribution of particles
across tree pieces and eventually precludes a good balance of load across
processors. In such scenarios, it becomes useful to repeat the full domain
decomposition more frequently.

The first stage of domain decomposition, as described in
Section~\ref{sect:parapproach}, involves a series of histogramming steps to
determine a set of splitters that partition the simulation domain into
tree pieces of roughly uniform particle count. This is implemented in terms of
broadcasts from a single sorter object, which refines the splitters, to the
tree pieces, and reductions of particle counts for each bin back to the sorter
process. In strong scaling scenarios for highly clustered datasets, domain
decomposition may become a performance bottleneck, as the number of splitters
generally depends on the number of processors used in the run. As such, we
implemented a number of optimizations aimed at improving SFC domain
decomposition performance. First, we replaced the broadcast of SFC keys from the
sorter object with the broadcast of a bit vector indicating which of the bins
evaluated in the previous step need further refinement. From the bit vector, the
set of splitters to evaluate is determined once at each SMP node, and
delivered to all tree pieces at that node for evaluation. This optimization
greatly reduced the size of the buffers being broadcast. Secondly, we noticed
that some histogramming steps were much more expensive than others, due to
involving more splitters. This was particularly true for the first
and last steps. The first histogramming step involved a full set of splitters
due to none having been finalized yet. For this step, we were able to
remove the broadcast of splitters by having tree pieces reuse the splitters
determined the last time domain decomposition was done. We were also able to
eliminate the last histogramming step in the original algorithm, in which the
final set of splitters was broadcast to the tree pieces to collect a full
histogram of particle counts. Instead, we modified the sorter object to preserve
particle counts for all previously finalized splitters, so as to have the full
set of counts at the end.

These optimizations significantly improved domain decomposition performance.
For runs of the \emph{cosmo25} dataset on Blue Waters, the time for a full domain
decomposition was reduced from 3.22 s to 1.52 s on 1024 nodes, a speedup of 2.1.

\subsection{Hierarchical Multistep Load Balancer}
\label{sect:hlb}
Even if domain decomposition assigns almost equal number of particles to tree
pieces, density variations in different regions of the simulated space can
result in load imbalance. We experimented with domain decomposition based on
load but the basic approach was not ideal for multi-stepping simulations as it
led to large jumps in boundaries and significant movement of particles. Since
execution time is determined by the most loaded processor, it becomes important
to address the load imbalance problem without significant additional overhead.

Load balancing in Charm++ applications like \changa{} is normally achieved by
over-decomposing the problem into many more objects than processors and letting
the Charm++ dynamic load balancing framework balance the load by mapping the
objects to processors~\cite{GengbinThesis}.  The framework can automatically
instrument the computation load and communication pattern of tree pieces and
other objects and store it in a distributed database.  This information is then
used by the load balancing strategies, which we optimized for \changa{}, to map the
objects to processors. Once the decision has been made, the load balancing
framework migrates the objects to newly assigned processors.  Alternatively, the
load of the objects and their communication pattern can be determined using a
model based on a priori knowledge.  But for \changa{}, we find that determining the
load based on a heuristic called the \emph{principle of persistence} is more
accurate. Based on this heuristic we use recent history to determine the load of
near-future iterations. This scheme works well for single-stepping simulations at
a relatively small scale.  However, multi-stepping simulations at very large scale
impose several new challenges.

First, multi-stepped execution introduces some challenges in the
measurement based load balancing to obtain accurate load information. Substeps within a big step in a
multi-step run have selected number of active particles. Predicting the load of a
tree piece based on the preceding substep will result in discrepancy between the
expected load and the actual load. Therefore, we instrument and store the load
of the tree pieces for different substeps/rungs separately. Whenever particles
migrate from one tree piece to another, they carry a fraction of their load for
the corresponding rungs for which they were active and contribute that to the
new tree piece. This enables us to achieve very accurate prediction of
the load of a tree piece for each substep even with migrations and multi-stepping.

Secondly, it is very challenging to collect communication pattern information in
\changa{}, even at small core count, due to a very large number of messages in the
simulation, which may incur significant overhead on memory when performing load
balancing. 
Therefore, we used an alternate strategy to implicitly
take communication into account during load balancing by using an ORB-based
(Orthogonal Recursive Bipartitioning) strategy, which preserves the
communication locality.

Lastly, in extremely large scale simulations, load balancing itself becomes
a severe bottleneck. The original centralized load
balancing strategies, where load balancing decision is made on one central
processor, do not scale beyond a few hundred processors, which
makes them infeasible for large scale simulations. To overcome this challenge,
we implemented a
scalable load balancing strategy suitable for multi-stepped execution based on the hierarchical load balancing framework
~\cite{GengbinThesis,NamdSC11} in Charm++ runtime. This new load balancing strategy performs ORB to distribute
the tree pieces among processors. The processors are divided into independent
groups organized in hierarchical fashion. Each group consists of $512$
processors. At each level of the hierarchy, the root performs the load balancing
strategy for the processors in its sub-tree. We found that two levels of
hierarchy is enough to achieve good load balance with little overhead. 
At higher levels of the hierarchy refinement based load balancing strategy,
which minimizes the migration by considering the current assignment of tasks, is
used. At the lowest level of the hierarchy we use ORB to partition the
tree pieces among the processors in that sub-group. 
The load balancer collects
the centroid information of tree pieces along with their load. Taking the
centroids into account, the tree pieces are spatially partitioned into two sets
along the longest dimension. Similarly, at each stage of partitioning, the
processors are also partitioned. During partitioning, tree pieces
are divided into two partitions such that the
loads of the partitions are almost equal. This is done recursively until one
processor remains which is assigned the corresponding partition containing the
tree pieces.

Another optimization to further reduce the overhead of load balancing is 
to combine the node level global load balancing with the intra-node load
balancing strategies
described in Section~\ref{sect:optgrav}. We implemented such a two-level load balancing strategy, 
where the load is first balanced across SMP nodes, and then balanced 
inside each SMP node. The
ORB algorithm described above is done for nodes rather than processors. Once the
tree pieces are assigned to SMP nodes, they are further distributed among the PEs
in the SMP node using a \emph{greedy} strategy. This ensures that load is equally
distributed among the SMP nodes. We perform an additional step of refinement to
further improve the load balance for the rare cases when the load is not evenly
balanced.


\subsection{Performance Evaluation}
\begin{figure*}[t]
\centering
\begin{tabular}{@{}c@{}c@{}}
  \subfloat[Single Stepping] {
  \includegraphics[width=0.47\textwidth, trim=0 0 0 10]{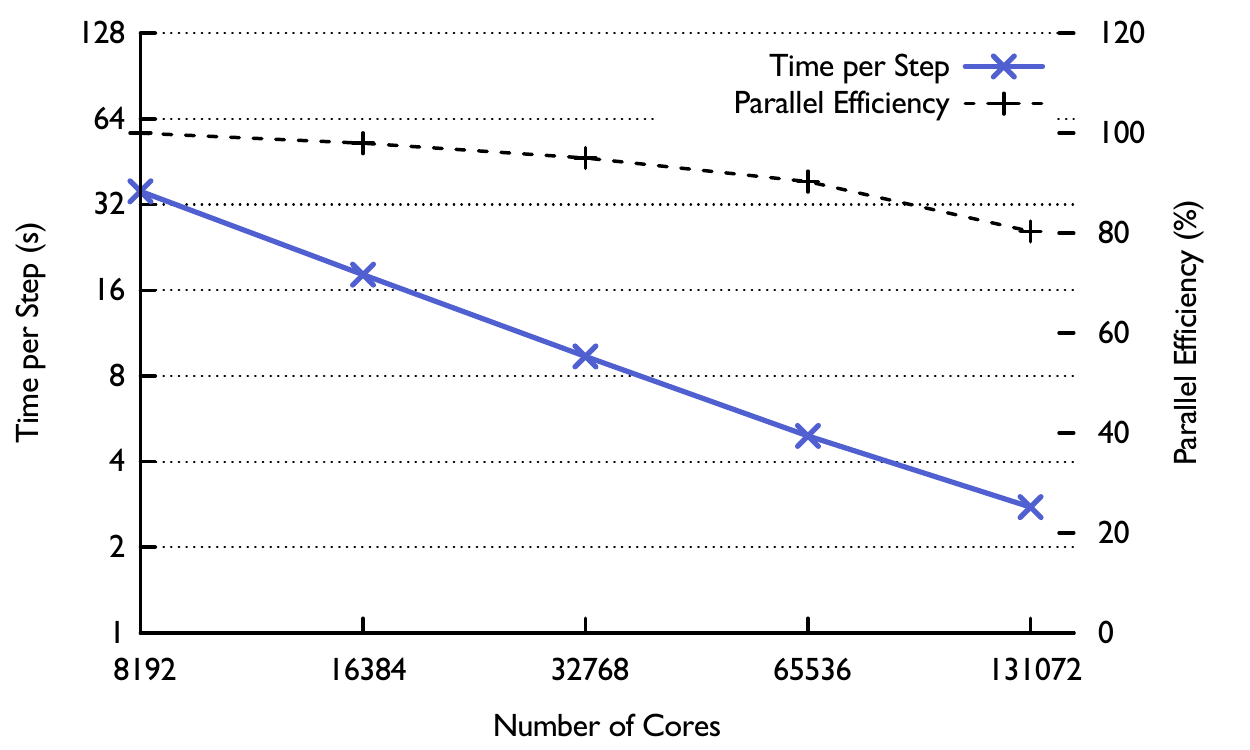}
  \label{fig:bw2bss}
  } &
  \subfloat[Multi Stepping] {
  \includegraphics[width=0.47\textwidth, trim=0 0 0 10]{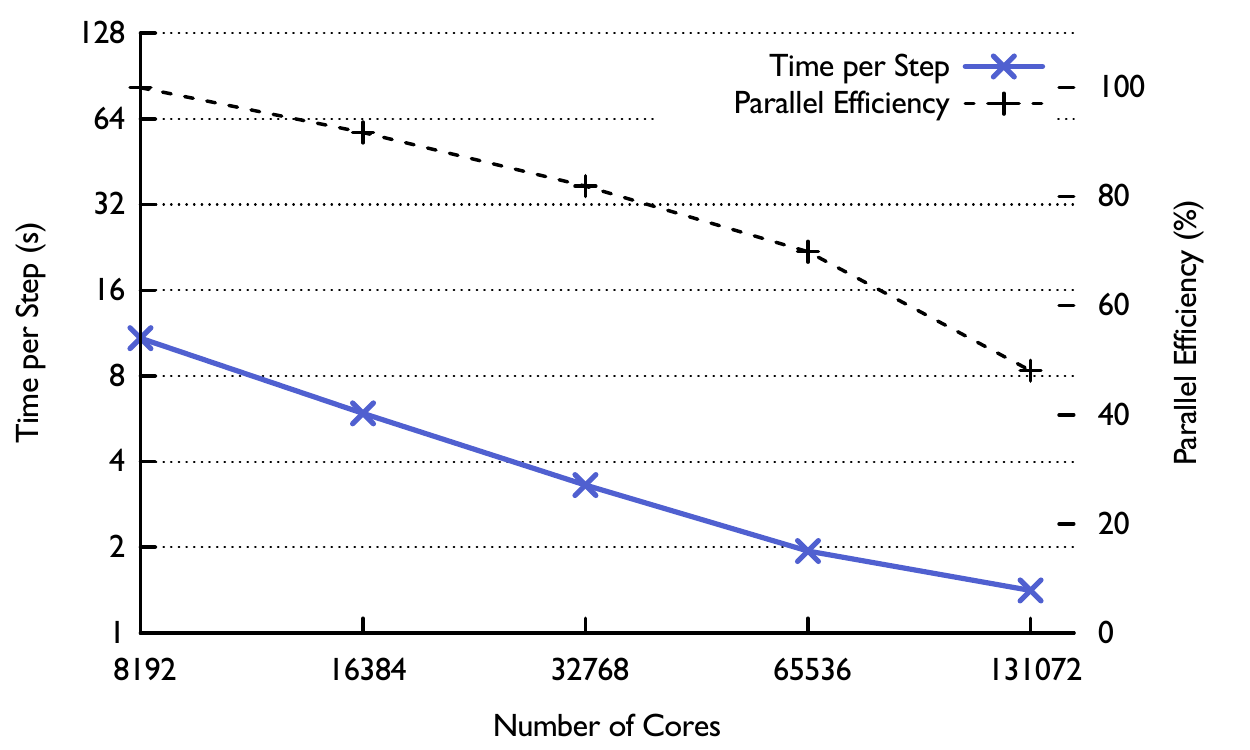}
  \label{fig:bw2bms}
  } 
\end{tabular}
\caption{Time per step and parallel efficiency for \emph{cosmo25} dataset on Blue Waters.}
\label{fig:bw2b}
\end{figure*}

\begin{table}[t]
\newcolumntype{C}{>{\centering\arraybackslash}X}
\begin{tabularx}{\columnwidth}{|C|C|C|C|C|C|C|}
\hline
{\bf \#cores} & {\bf Gravity} & {\bf DD} & {\bf TB} & {\bf LB} & {\bf Step time} & {\bf 16 Step time} \\ 
\hline
8192 & 33.433 &  0.441 & 0.292 & 1.423 & 35.589 & 569.424 \\
16384 & 16.952 &  0.210  & 0.148 & 0.851 & 18.161 & 290.576 \\
32768 & 8.643 & 0.132 & 0.091 & 0.496 & 9.362 & 149.792 \\
65536 & 4.395 & 0.163 & 0.073 & 0.295 & 4.926 & 78.816 \\
131072  & 2.353 & 0.134 & 0.066 & 0.216 & 2.769 & 44.304 \\
\hline
\hline
8192 & 7.45  & 0.83 &  0.47 &  2.1 & 10.85 & 173.6 \\
16384 & 3.73 &  0.79 &  0.32 &  1.07 &  5.91 & 94.56 \\
32768 & 2.1 & 0.46 &  0.2 & 0.55 &  3.31 & 52.96 \\
65536 & 1.1 & 0.35 &  0.12 &  0.37 &  1.94 & 31.04 \\
131072 &  0.77 &  0.24 &  0.07 &  0.33 &  1.41 & 22.56 \\
\hline
\end{tabularx}
\caption{Breakdown of time for 1 step in seconds for \emph{cosmo25} dataset with
single-stepping (top half) and multi-stepping (bottom half) on Blue Waters}
\label{tab:2bssbw}
\end{table}

We now present the scaling performance of the \emph{cosmo25} simulation.
Figure~\ref{fig:bw2bss} shows the average time per iteration for this simulation
with single-stepping and figure~\ref{fig:bw2bms} shows the average time per
iteration with multi-stepping.  In a
multi-stepping run, $16$ substeps constitute a big step. To compare the time for
single-stepping and multi-stepping, a single big
multi-step covers the same dynamical time as 16 single steps. Table~\ref{tab:2bssbw} gives a break down of the
time taken for different phases for single-stepping and multi-stepping.  We can
see that at $8K$ cores the single-stepping simulation takes more than $3$ times
the time taken by multi-stepping and at $128K$ it takes twice as long. Note that
the gravity time for multi-stepping is $4.5$ times faster than single
stepping. Due to sufficient sequential work to overlap communication and
relatively balanced tree pieces, we are able to achieve $80\%$ efficiency for
single-stepping at $128K$ cores with an average step time of $2.7$ seconds. As
described in section~\ref{sect:optgrav} the multi-stepping run has many challenges
due to irregular distribution of particles in faster rungs. Incorporating the
improvements mentioned above,
we are able to scale to
$128K$ cores with an efficiency of $48\%$ with respect to $8K$ cores with
a time step of $1.4$ seconds. Note that if we consider the gravity force
calculation time, we achieve an efficiency of $60\%$ and the gravity time is $3$
times faster in multi-stepping in comparison to the single-stepping run.

%


\section{Conclusion}
\label{sect:conc}
In this paper, we have described the design and features of our highly scalable
parallel gravity code \changa{} and went into the details of scaling challenges for
clustered multiple time-stepping datasets. We have presented strong scaling
results for uniform datasets on up to $512K$ cores on Blue Waters evolving $12$
and $24$ billion particles. We also present strong scaling results for
\emph{cosmo25} and \emph{dwarf} datasets, which are more challenging due to
their highly clustered nature. We obtain good performance on up to $128K$ cores
of Blue Waters and also show up to a $3$ fold improvement in time with multi-stepping over
single-stepping.  

Many features of the \charm{} runtime system were used to achieve
these results.  Starting with the standard load balancing and overlap of
communication and computation enabled by the over-decomposition
strategy, we employed a number of \charm{}'s features.  Of particular
importance were features that allowed us to replace parts of our
algorithm that scaled as the number of cores, such as quiescence
detection for particle movement and the hierarchical load balancer.
Also of importance were features such as CkLoop, SMP Cache and node
level load balancing, that exploited SMP features of almost
all modern supercomputers.  With these features, we can bring to bear
the computational resources of many 100s of thousands of processor
cores on the highly clustered, large dynamic range simulations that
are necessary for understanding the formation of galaxies in the
context of of large scale structure.

\begin{backmatter}

\section*{Author's contributions}
T.Q. is the primary researcher and supervisor of ChaNGa project. T.Q., P.J.,
along with various contributors developed the code.  T.Q. and F.G., and others
verified the code for cosmological simulations. H.M., L.W., T.Q. and L.K. came
up with the techniques mentioned in the paper for scaling the application. H.M.
developed the dynamic load balancing techniques and optimizations for the
various phases of the simulation. G.Z. and H.M. developed the hierarchical load
balancer. L.W. worked on the domain decomposition optimizations. P.J. developed the
SMP cache optimization. G.Z. optimized the performance for the Blue Waters
hardware. H.M. performed the scaling experiments with the help
from T.Q., G.Z. and L.W.  All the authors discussed the results and contributed
extensively to the writing of the paper.

\section*{Acknowledgements}
\label{sect:ack}
\changa{} was initially developed under NSF ITR award 0205413.
Contributors to the development of the code include Graeme
Lufkin, Sayantan Chakravorty, Amit Sharma, and Filippo Gioachin.
This research is part of the Blue Waters sustained-petascale computing project,
which is supported by the National Science Foundation (award number OCI
07-25070) and the state of Illinois.
HM and LW was supported by NSF award AST-1312913.
TQ and FG where supported by NSF award AST-1311956.
Use of Bluewaters was supported by NSF PRAC Award 1144357.
We made use of pynbody (https://github.com/pynbody/pynbody) to create
figure \ref{fig:blob_changa}, and we thank Andrew Pontzen for
assistance in creating that figure.

\theendnotes

\bibliographystyle{bmc-mathphys} 
\bibliography{cited,group,cosmo}

\end{backmatter}

\end{document}